 \documentclass{cambridge6A}
  \usepackage{graphicx}

  \usepackage{amsmath,amssymb}
  
\newcommand{\pd}[2]{\frac{\partial #1}{\partial #2}}
  
\begin{document}

\renewcommand*\thesection{\arabic{section}}
\renewcommand*\thesubsection{\arabic{section}.\arabic{subsection}}

\setcounter{equation}{0}
\numberwithin{equation}{section}
\setcounter{figure}{0}
\renewcommand{\thefigure}{\arabic{figure}.}

\chapterauthor{Toby Wiseman\\ ~ \\
\textit{Theoretical Physics Group, Blackett Laboratory, Imperial College, London SW7 2AZ, U.K.}}

\chapter*{Numerical construction of static and stationary black holes}

\begin{abstract}\small
This book chapter gives an introductory review of a numerical framework for finding static and stationary vacuum black hole solutions. Such methods may be applied to explore the exotic space of higher dimensional black holes that are thought to exist.
\end{abstract}

\copyrightline{Chapter of the book \textit{Black Holes in Higher Dimensions} to
be published by Cambridge University Press (editor: G. Horowitz)}

\section{Introduction}

Whilst the black holes of 4D are well mannered, being spherically symmetric or having special algebraic properties which enables them to be found analytically, moving beyond 4D many solutions of interest appear to have no manners whatsoever.
The problem of finding these unruly black holes becomes that of solving the non-linear coupled set of partial differential equations (PDEs) for the metric components given by the Einstein equations. In general it is unlikely that closed form analytic solutions will be found for many of the exotic black holes discussed earlier in this book. If we are to understand their properties then we must turn to numerical techniques to tackle these PDEs. It is the purpose of this chapter to develop general numerical methods to address the problem of finding static and stationary black holes. 

Surely the phrase `the devil is in the details' could not have a truer application than to numerics. The emphasis of this chapter will be to provide a road map to tackling the problem, where we formulate the problem in as unified, elegant and geometric a way as possible. We will also discuss concrete algorithms to solve the resulting problem, but the extensive details of implementation will not be addressed, probably much to the readers relief. Such details can be found in the various articles cited in this chapter. 

Given the extensive efforts that have gone into understanding dynamical numerical simulation of gravity, the most obvious approach to find static and stationary solutions is to simulate a dynamical collapse of matter that is likely to form a solution of interest, or alternatively to simulate vacuum gravity starting with initial data that appears to already contain the black hole. 
Such a dynamical approach is indeed possible but is not the approach we will develop here for several reasons. 
Firstly, whilst such dynamical evolutions are well understood they are very complicated, typically requiring large computer resources. 
An important point is that in order to find an accurate stationary solution one would have to run a dynamical simulation for a long time to ensure that the resulting solution has indeed settled down, losing all its excitations to gravity waves, and this can be a serious challenge.
Secondly, since one necessarily would be following the detailed evolution to form a horizon, to watch it radiate gravity waves and ring down, one is certainly doing far more work than is necessary if all that is required is the final stationary solution. 
Thirdly, many of the families of exotic solutions that have been discussed are likely to be unstable, at least for certain ranges of parameters. It is still interesting to find such black holes in order to elucidate the global structure of the space of solutions. For example in Chapter 4 we discussed that in Kaluza-Klein theory the inhomogeneous black strings and some of the localised black holes are thought to be unstable. However to confirm the elegant picture that these solutions are continuously connected to each other requires us to find these unstable solutions.
In principle one might try to tune initial data to find an unstable solution, although in practice this is likely to be very hard.
We also mention a more speculative concern, namely that in higher dimensions it appears that cosmic censorship does not hold \cite{Lehner:2010pn}, and therefore it is unclear how generically one might expect to encounter singularity formation in dynamics.

For all these reasons here we focus on  finding static or stationary black holes by solving the static or stationary Einstein equations directly. Since these are PDEs the most important question is what character they have, as this will determine how to approach the problem. In the context of dynamics the Einstein equations should be thought of as having hyperbolic character i.e. on small scales one has wave propagation along a light-cone. Consequently the dynamical Einstein equations are thought of as an initial value problem, with data being specified on a Cauchy surface or past light cone. 
In contrast to this the static and stationary problem should be thought of as having elliptic character. One solves elliptic systems as boundary value problems where for a second order elliptic system one piece of data, for example Dirichlet/Neumann/Robin/oblique, is given on all boundaries. Physically in our static or stationary context such boundary conditions will correspond to ensuring horizon regularity whilst also prescribing some particular asymptotic behaviour.

Our focus is to present the problem in an elliptic framework where one requires only entirely conventional techniques and modest desktop computing resources. In particular we will discuss standard methods to solve the elliptic systems, namely relaxation and Newton's method. In addition the approach is amenable to the full variety of methods to represent the solutions, for example the classical finite difference approach, or more modern spectral, pseudo-spectral and finite element methods.

Before embarking on our discussion of the numerical stationary problem for higher dimensional black holes some comments on the history of this problem are in order. 
Only with the rather recent revelation that black hole uniqueness breaks down in dimensions $D>4$ did numerical exploration begin of gravity in higher dimensions. Of course in 4 dimensions uniqueness of Kerr meant that numerical work was not traditionally directed at vacuum black holes. 
Indeed a key ingredient of the classical proofs of uniqueness was to formulate the 4D stationary axisymmetric vacuum problem as an elliptic system \cite{Carter}.
However the numerical stationary elliptic problem has featured prominantly in the context of relativistic stars, particularly in the axisymmetric context, and there is a distinguished history of numerical work in this area beginning in the early 1970's (see for example the seminal works \cite{W72, BS74, BI76}). 
For a review of this fascinating field the reader is referred to the Living Review article \cite{Stergioulas}. 
In the context of non-vacuum static and stationary black holes such numerical methods were applied in \cite{Kleihaus:1997ic} and \cite{Kleihaus:2000kg} to find exotic 4D charged solutions.
In all these cases the approaches developed are generally based on the Weyl-Papapetrou form of the metric, exploiting the fact that the metric only depends non-trivially on two coordinates and has two commuting Killing vectors, so that solving the  Einstein equations is consistently reduced to an elliptic problem. 
In the analytic context recent uniqueness theorems for vacuum stationary black holes in higher dimensions \cite{Morisawa:2004tc,Hollands:2007aj} also phrase the problem as an elliptic system, and so far have restricted attention to D dimensional spacetimes with D-2 commuting Killing vectors, where again the Weyl-Papapetrou form is used (see also \cite{Harmark:2004rm}).

Numerical methods for metrics depending non-trivally on two coordinates, but without the restriction of having D-2 commuting Killing vectors were given in  \cite{star, Wiseman:2002zc} and applied to higher dimensional black holes. An analog of the Weyl form was employed for the metric. It was shown that a subset of the Einstein equations are elliptic in the metric components and may be solved  regarding the remaining Einstein equations as constraints.  One must show these constraints may be consistently solved by consideration of the various boundary conditions. Being based on this analog Weyl form, these methods are manifestly non-covariant and by construction can only be applied to problems depending non-trivially on two coordinates. 
Whilst they have yielded interesting results they give a rather unstable numerical scheme for solutions in $D>4$ which have axes of rotational symmetry (such as localised Kaluza-Klein black holes), presumably due to the lack of covariance, and are therefore hard to use in practice. 
We emphasize that the methods we will develop here will be fully covariant, and may be applied to problems depending non-trivially on an arbitrary number of coordinates. In our experience they are much better behaved in practice, for example, having no difficulties with axes of symmetry.

For simplicity this chapter will focus entirely on finding vacuum black holes with no cosmological term and with a  single component non-extremal Killing horizon. We will spend considerable time treating the case of static black holes, where most previous work has been directed. Indeed the numerical solutions of \cite{HeadrickKitchenTW} discussed in Chapter 4 where found precisely using the methods we discuss.
 By analytic continuation of time this problem has an elegant geometric formulation as finding Ricci flat Riemannian solutions. We will study in detail the issue of formulation as an elliptic boundary value problem and give two algorithms to solve the resulting PDEs. In the remainder of the chapter we will show that the stationary case, which must directly be treated in Lorentzian signature, can also be thought of as an elliptic boundary value problem.

\section{Static vacuum black holes}

In this section we treat the static vacuum case following the approach of \cite{HeadrickKitchenTW} with Headrick and Kitchen.
Let us consider a general non-extremal static black hole solution with a single component horizon so that we may write the metric as,
\begin{eqnarray}
ds^2 = - N(x)^2 dt^2 + h_{ij}(x) dx^i dx^j
\end{eqnarray}
where $\partial / \partial t$ is the static timelike Killing vector, and at the horizon the norm of this vector vanishes so $N = 0$. The zeroth law implies that the surface gravity given by the function $\kappa = \partial_n N |_{N=0}$, where $n$ is the unit normal vector to the horizon in a constant $t$ slice, is actually a constant.  A standard result of Euclidean quantum gravity is that any such static black hole may be analytically continued to imaginary time $\tau = i t$ to yield a Riemannian manifold with metric, 
\begin{eqnarray}
ds^2 = + N(x)^2 d\tau^2 + h_{ij}(x) dx^i dx^j
\end{eqnarray}
where upon making $\tau$ an angular coordinate with period $\tau \sim \tau + 2 \pi / \kappa$ the metric at the horizon is smooth with no boundary there. To manifest this we may take Gaussian normal coordinates to the horizon where $x^i = \{ r, x^a \}$ and the horizon is located at $r = 0$. Then near the horizon the metric goes as,
\begin{eqnarray}
ds^2 &\sim& \left( \kappa^2 r^2 d\tau^2 + dr^2 \right) + \tilde{h}_{ab}(r, x) dx^a dx^b
\end{eqnarray}
and we see that the shrinking Euclidean time circle forms the angle of polar coordinates in $\mathbb{R}^2$, with $r$ the radial coordinate. Whilst this polar coordinate system breaks down at the origin $r = 0$, simply by taking `Cartesian' coordinates $X = r \cos{\kappa \tau}$ and $Y = r \sin{\kappa \tau}$, one can write the metric in a chart that covers the horizon. 

Thus a static black hole can be written as a smooth Euclidean geometry, where Euclidean time $\tau$ is periodic, and $\partial  / \partial \tau$ generates a $U(1)$ isometry. There is no boundary at the horizon and the geometry is perfectly smooth there. The horizon is marked by the vanishing of the vector $\partial  / \partial \tau$, and consequently the horizon forms the fixed point set of the static isometry.

Solving the vacuum Einstein equation without cosmological term is equivalent to finding a geometry that is Ricci flat, so $R_{\mu\nu} = 0$. Finding a vacuum static black hole solution can then be viewed as part of the more general problem of
finding Ricci flat Riemannian geometries, the only special feature of a static black hole geometry being the $U(1)$ isometry, generated by the hyper-surface orthogonal vector $\partial / \partial \tau$, which leaves a codimension two submanifold (the horizon) fixed. 

There are several attractive features of this way of thinking. Firstly  there is in principle  no boundary at the horizon, and from the point of view of our boundary value problem, no boundary condition to impose there. The `in principle' qualification, which we will explain in more depth later, refers to the fact that if one wishes to use coordinates that manifest the $U(1)$ isometry then these will not cover the horizon, in exactly the same way as polar coordinates do not cover the origin point. Whilst in practice it is sensible to adapt coordinates to any isometries and we shall deal with this in detail later, for the time being let us take a more formal view ignoring issues of implementation such as which coordinates to take. In principle one can always take coordinates that are regular at the horizon (the `Cartesian' coordinates above) and then there is no boundary.

The second attractive feature is that the Euclidean continuation is precisely what is done in semi-classical quantum gravity to consider the canonical ensemble i.e. to work at fixed finite temperature. In particular the proper size of the Euclidean time circle asymptotically has the interpretation of being the inverse temperature of the black hole. Far from the black hole horizon we will impose boundary conditions on the metric that involve fixing amongst other things this proper size of the time circle. This boundary value formulation then naturally leads us to fix a very physical quantity, the temperature.

Thirdly, the problem of finding Ricci flat Riemannian metrics is one with wider application than just finding static black holes. For example, numerically finding the exotic Calabi-Yau geometries \cite{CY,Douglas:2006hz} that underpin certain string theory compactifications is currently required to make detailed predictions of low energy phenomenology. In addition, particularly in the case of such K\"ahler metrics  there may be new ways to think about the problem inspired by the geometry \cite{Donaldson,Headrick:2009jz} which might in the future be useful in our black hole context.

Let us now consider how to formulate the problem of finding Ricci flat Riemannian geometries as an elliptic boundary value problem.

\subsection{The Harmonic Einstein equation}
\label{sec:harmeqn}

The vacuum Einstein equation $R_{\mu\nu} = 0$ is a second order quasi-linear PDE in the metric components. Perturb the metric about some background $g_{\mu\nu}$ by a perturbation $h_{\mu\nu}$, and then,
\begin{equation}
\label{eq:linRicci}
\delta R_{\mu\nu} \equiv \Delta_R h_{\mu\nu} = \Delta_Lh_{\mu\nu} + \nabla_{(\mu}v_{\nu)}\,,
\end{equation}
where
\begin{equation}
\Delta_Lh_{\mu\nu} \equiv 
-\frac12\nabla^2h_{\mu\nu} - R_\mu{}^\kappa{}_\nu{}^\lambda h_{\kappa\lambda} + {R_{(\mu}}^\kappa h_{\nu)\kappa} \,,\qquad v_\mu \equiv \nabla_\nu {h^\nu}_\mu - \frac12\partial_\mu h\,,
\end{equation}
and $\Delta_L$ is the usual Lichnerowicz operator. The principle part of $\Delta_R$, which we denote $P_g$, is given locally by taking only the two derivative terms,
\begin{equation}
P_g h_{\mu\nu} = \frac{1}{2} \left( g^{\alpha \beta} \partial_\mu \partial_\alpha h_{\beta \nu} +  g^{\alpha \beta} \partial_\nu \partial_\alpha h_{\beta \mu} - g^{\alpha\beta} \partial_\alpha \partial_\beta h_{\mu\nu} - g^{\alpha\beta} \partial_\mu  \partial_\nu h_{\alpha\beta} \right)
\end{equation}
and this linear operator controls the very short wavelength behaviour of perturbations which determines the character of the equations $R_{\mu\nu}$ about the background $g$. The condition that $R_{\mu\nu} = 0$ is elliptic about some background $g$ gives the requirement that if one takes $h_{\mu\nu} = a_{\mu\nu} e^{i k_\alpha x^\alpha}$ for some constants $a_{\mu\nu}$ and any real non-zero $k_\mu$, then $P_g  h_{\mu\nu} \ne 0$ everywhere. Physically this condition means that nowhere can we find a point where short wavelength perturbations in a particular direction propagate as a wave.

We see that for perturbations of the form $h_{\mu\nu} = \partial_{(\mu} u_{\nu)}$, where $u$ is some vector field, then $P_g h_{\mu\nu} = 0$. Thus the Ricci flatness condition $R_{\mu\nu} = 0$ is not an elliptic equation.
Such a perturbation can be thought of as a short wavelength infinitesimal diffeomorphism generated by $u$. This diffeomorphism is given by $h_{\mu\nu} = \nabla_{(\mu} u_{\nu)}$ but for a $u$ which varies on very short scales, then $ \nabla_{(\mu} u_{\nu)} \sim \partial_{(\mu} u_{\nu)}$. Hence we may see lack of ellipticity of $R_{\mu\nu} = 0$ as a consequence of gauge invariance. Without an elliptic set of PDEs we cannot treat the system as a boundary value problem as we wish to. 
In order to proceed we must therefore break this gauge invariance. However lifting gauge invariance need not imply breaking covariance. We emphasise that the method described below will indeed be fully covariant.

Instead of considering the vacuum Einstein equation $R_{\mu\nu} = 0$, we will consider what we term the \emph{Harmonic Einstein} equation,\footnote{Note that we have referred to this equation also as the Einstein-DeTurck equation in the work \cite{HeadrickKitchenTW,PFLuciettiTW}. } 
$R^H_{\mu\nu} = 0$ where,
\begin{equation}
\label{eq:DeTurck}
R^H_{\mu\nu} \equiv R_{\mu\nu} - \nabla_{(\mu} \xi_{\nu)} \, , \quad \xi^\alpha \equiv g^{\mu\nu} \left( \Gamma^{\alpha}_{~\mu\nu} - \bar{\Gamma}^{\alpha}_{~\mu\nu} \right) \, .
\end{equation}
$\Gamma$ is our usual Levi-Civita connection of $g$, and $\bar{\Gamma}$ is another connection which we are free to choose, and then consider fixed. We term $\bar{\Gamma}$ the reference connection. Being constructed from the difference of two connections, the quantity $\xi$ is a globally defined vector field. 
The equations $R^H_{\mu\nu} = 0$ have the great virtue of being elliptic. The principle part of the linearisation about a background $g$ is simply,
\begin{equation}
\label{eq:symbol}
P^H_g h_{\mu\nu} = - \frac{1}{2} g^{\alpha\beta} \partial_\alpha \partial_\beta h_{\mu\nu} 
\end{equation}
and thus for \emph{any} Riemannian background $g$ this clearly is elliptic. Taking $h_{\mu\nu} = a_{\mu\nu} e^{i k_\alpha x^\alpha}$, then $P^H_g  h_{\mu\nu} = a_{\mu\nu} k^\alpha k_\alpha$ which indeed only vanishes for vanishing $h_{\mu\nu}$ or $k_\mu$ as required for ellipticity. An important point is that we have used the fact that we have analytically continued to Euclidean signature. For a Lorentzian signature metric $g$, then $P^H_g  h_{\mu\nu} = 0$ at a point $x$ if we pick any non-zero but null vector $k$. In this Lorentzian case the Harmonic Einstein equation has hyperbolic character.

The Harmonic Einstein equation has been used in the context of analysis for decades for both the Riemannian elliptic problem as well as the Lorentzian dynamical hyperbolic problem (see for example \cite{Bruhat}), with variations in the precise definition of the vector that all lead to elliptic/hyperbolic equations at least in the neighbourhood of a Ricci flat solution. 

The choice of vector field we have taken above in \eqref{eq:DeTurck} is due to DeTurck who introduced it in the Riemannian context and later used it to show Ricci flow is parabolic as we discuss later  \cite{DeTurck}. 
For simplicity in what follows we reduce the freedom in the definition above and will take $\bar{\Gamma}$ to be the Levi-Civita connection of a reference metric $\bar{g}$ which we are free to choose and then consider fixed \cite{HawkingEllis}. In this case we may write,
\begin{eqnarray}
\label{eq:xi}
\xi_\mu =  g^{\alpha\beta} \left( \bar{\nabla}_{(\alpha} g_{\beta)\mu} - \frac{1}{2} \bar{\nabla}_\mu g_{\alpha\beta} \right) \, ,
\end{eqnarray}
where $\bar{\nabla}$ is the covariant derivative of the metric $\bar{g}$.
\footnote{We note this is close to, although not the same as the Bianchi choice of vector field used, for example, in \cite{Anderson}. In particular the Bianchi choice does not lead to the simple principle symbol \eqref{eq:symbol} except for metrics $g$ close to the reference metric $\bar{g}$.  }

In D dimensions there are D local coordinate degrees of freedom to fix in order to lift the gauge invariance of the Ricci flatness condition. The condition $\xi^\mu = 0$ precisely provides these D additional local conditions. The DeTurck choice of $\xi$ can be thought of as a global version of the generalised harmonic coordinates introduced by Friedrich in the Lorentzian hyperbolic context and independently employed later by Garfinkle for numerical evolutions \cite{Friedrich,Garfinkle:2001ni} and used extensively since then. 
In generalised harmonic coordinates one writes $\xi^\alpha = g^{\mu\nu} \Gamma_{~\mu\nu}^{\alpha} + H^\alpha$ in local coordinates for some choice of $H^\alpha$, although we note that $H^\alpha$ is not a vector field globally. Locally this is the same as the definition above, where $H^\alpha = - g^{\mu\nu} \bar{\Gamma}_{~\mu\nu}^{\alpha}$. 
The vanishing of $\xi^\mu$ can be thought of as a generalised harmonic gauge condition.
\footnote{Note the subtle difference between the usual use of generalised harmonic coordinates, where one locally fixes $H$, and the DeTurck case here where instead one fixes $\bar{\Gamma}$. These are inequivalent as the relation between them involves the metric $g$.}
Given a chart the coordinates $x^\alpha$ are functions over the part of the manifold covered by the chart. If in this chart we took $\bar{\Gamma}^\alpha_{~\mu\nu} = 0$, then vanishing $\xi$ would imply $\nabla_S^2 x^\alpha = 0$ where $\nabla_S^2$ is the scalar Laplacian, and hence the local coordinates would be harmonic functions - so called harmonic coordinates. For a general choice of reference connection, we have $\nabla_S^2 x^\alpha = H^\alpha$ corresponding to generalised harmonic coordinates. 
From now on we shall refer to a metric with $\xi = 0$ for our DeTurck choice of $\xi$ in \eqref{eq:DeTurck} as being in the generalised harmonic gauge.
\footnote{
Since $\xi = 0$ can be viewed as elliptic equations for the coordinate functions we see that whilst it is able to constrain the local degrees of freedom in the gauge, as with any elliptic equation one must specify boundary data, and this global gauge freedom must be fixed.
When one specifies charts, one must specify the domain of these charts in $\mathbb{R}^D$ and this data precisely gives Dirichlet boundary conditions for the harmonic coordinate functions of the chart. This Dirichlet data for all the coordinate functions then implies all global data specifying the gauge is used up. Thus in the elliptic context harmonic coordinates are convenient as you have precisely the freedom to choose fixed charts on the manifold, and having made this choice there is no gauge freedom left. 
If the manifold has a boundary (or fictitious boundary as we discuss later) one may freely use a chart adapted to this boundary, so that the boundary is at some constant coordinate location, without being concerned that the boundary position in the harmonic coordinates might be something one must solve for as in some other gauge choices.
}

A Ricci flat solution in generalised harmonic gauge $\xi = 0$ does solve the Harmonic Einstein equation $R^H_{\mu\nu} = 0$. However the careful reader will notice that whilst the PDEs $R^H_{\mu\nu} = 0$ are indeed elliptic for a Riemannian $g$ (or indeed hyperbolic for Lorentzian $g$) there is no reason to suspect a priori that a solution to $R^H_{\mu\nu} = 0$ has anything to do with a solution to the Ricci flatness condition.

We now consider why solving $R^H_{\mu\nu} = 0$ might lead to a Ricci flat solution presented in generalised harmonic coordinates. The situation is simplest in the Lorentzian hyperbolic context where 
the answer lies in the contracted Bianchi identity applied to the Harmonic Einstein equation. In either signature this yields the linear PDE
\begin{eqnarray}
\label{eq:Bianchi}
\nabla^2 \xi_\mu + R_{\mu}^{~\nu} \xi_\nu = 0 
\end{eqnarray}
for the vector $\xi^\mu$. If one ensures $\xi^\mu$ and its normal derivative vanish on a Cauchy surface, then since the linear equation above is a wave equation, i.e. hyperbolic, for Lorentzian $g$ then $\xi^\mu$ must remain zero under evolution of the metric in time. Hence in the dynamical hyperbolic context one simply imposes the vanishing of $\xi$ and its time derivative as constraints on the initial data for the metric, then solves the hyperbolic Harmonic Einstein equation, and is guaranteed to recover a solution of the actual Einstein equation $R_{\mu\nu} = 0$ in coordinates defined by $\xi^\mu = 0$. 

In the Riemannian elliptic context we are interested in the situation is a little more complicated. As we shall see later we must supply data on any boundaries, give an initial guess and essentially hope for the best. Our initial guess and our reference metric will typically be far from a Ricci flat solution. Hence we cannot consider the Harmonic Einstein equation only in the neighbourhood of a Ricci flat solution, as the entire problem is to find that solution. 
One might then imagine that the situation is hopeless in the elliptic context, that solving $R^H_{\mu\nu} = 0$ is a quite unrelated problem to solving $R_{\mu\nu} = 0$. This is not the case as we now discuss.

\subsection{Ricci flat solutions and Ricci solitons}
\label{sec:solitons}

A solution to the Harmonic Einstein equation $R_{\mu\nu} =  \nabla_{(\mu} \xi_{\nu)}$ with non-vanishing $\xi$ is called a \emph{Ricci soliton}. Obviously we are interested in Ricci flat solutions rather than solitons. As we proceed to discuss, fortunately the existence of solitons is rather constrained provided we choose our boundary conditions appropriately.

Suppose we have boundaries or asymptotic regions in our problem and prescribe some data for the metric there, compatible with the ellipticity of the Harmonic Einstein equation. This data for the metric defines certain behaviour for the vector field $\xi$. 
Consider as an example a manifold with a boundary. Taking local coordinates near the boundary we may write the metric as,
\begin{eqnarray}
ds^2 = \alpha^2 dw^2 + \gamma_{ij} ( dx^i + \beta^i dw) ( dx^j + \beta^j dw) 
\end{eqnarray}
where the boundary is located at $w = 0$. Geometrically we might expect to try to fix some boundary condition involving the induced metric $\gamma_{ij}|_{w = 0}$ and the extrinsic curvature, $K_{ij} = \frac{1}{2 \alpha} \left( \partial_w \gamma_{ij} - 2 \nabla_{(i} \beta_{j)} \right)|_{w = 0}$ (where the covariant derivatives and metric contractions are taken with respect to $\gamma$). We may regard $\gamma_{ij}$ as the geometric Dirichlet data, and $K_{ij}$ which involves the normal derivative of $\gamma_{ij}$ as the Neumann data.
However $\gamma_{ij}$ and $K_{ij}$ only have $D(D-1)/2$ components, whereas the full metric $g_{\mu\nu}$ has $D(D+1)/2$. The Harmonic Einstein equation requires elliptic data for all these components of $g_{\mu\nu}$. How should we fix the remaining $D$ conditions? Since we are interested in Ricci flat solutions of $R^H_{\mu\nu} = 0$ where $\xi^\mu = 0$ we had better enforce this. The additional conditions that $\xi^\mu = 0$ precisely give the extra $D$ conditions for the metric components.\footnote{
Very interestingly Anderson has shown that one cannot impose $\xi = 0$ and require Dirichlet or Neumann boundary conditions (i.e. fixed induced metric or extrinsic curvature) as these are not well posed boundary conditions. Instead one may fix the conformal class of $\gamma_{ij}$ and the trace of $K_{ij}$ together with $\xi = 0$.\cite{Anderson}
} 

For the black holes discussed in this book we are not concerned with manifolds with boundary, but rather with asymptotic conditions. As we shall see explicitly later, we may impose an asymptotically flat or Kaluza-Klein condition also so as to ensure that $\xi \to 0$ at infinity. However this boundary example nicely illustrates that in order to give data for the Harmonic Einstein equation one should ensure that the data is consistent not only with imposing the geometric condition of interest, but also ensuring that $\xi$ vanishes. Obviously if one took boundary conditions that did not allow $\xi$ to vanish then generally one might still be able to solve $R^H_{\mu\nu} = 0$ but should not expect to find Ricci flat solutions but only solitons. 

Consider a metric $g$ and boundary/asymptotic conditions that we impose on it consistent with ellipticity. These boundary conditions impose certain behaviours on $\xi$ - for example $\xi = 0$ at a boundary as in the example above, or $\xi \to 0$ in an asymptotically flat or Kaluza-Klein region.
Then consider the linear vector operator,
\begin{eqnarray}
\label{eq:bianchi2}
\mathcal{D}_\mu^{~\nu} \equiv \nabla^2 \delta_\mu^{~\nu} + R_{\mu}^{~\nu}
\end{eqnarray}
 on this background so that we may write the Bianchi identity \eqref{eq:Bianchi} as $\mathcal{D} \cdot \xi = 0$.
 Take a vector field $\chi$ which has the same boundary or asymptotic behaviour as $\xi$ - for example, $\chi = 0$ on the boundary  described above, or $\chi \to 0$ in an asymptotically flat/Kaluza-Klein region. 
In order to find Ricci flat solutions to the Harmonic Einstein equation one should ensure that the metric boundary conditions lead to conditions on $\chi$ such that the linear elliptic vector problem $\mathcal{D} \cdot \chi = 0$ is well posed and admits the trivial solution $\chi = 0$.

Take the example of an asymptotically flat or Kaluza-Klein black hole. In the Riemannian signature with periodic time there are no boundaries and only the asymptotic region. The problem $\mathcal{D} \cdot \chi = 0$ with the condition $\chi \to 0$ asymptotically is indeed well posed and $\chi = 0$ is obviously a solution.

Note that we do not necessarily require the boundary condition that $\xi \to 0$ on all boundaries. For example, at a `fictitious boundary' (see later section \ref{sec:fictitious}) there are conditions as we shall see later where the tangential components of $\xi$ have vanishing normal derivative (a Neumann condition), rather than being forced to vanish. This still gives a well posed problem $\mathcal{D} \cdot \chi = 0$ consistent with a trivial solution.

Now that we have boundary conditions consistent with obtaining a Ricci flat solution, we may use the Bianchi identity to control the existence of solitons.  A necessary condition for a soliton to exist is that $\mathcal{D} \cdot \chi = 0$ must admit a non trivial solution for $\chi$. Alternatively we may say that the vector operator $\mathcal{D}$ with the boundary conditions given by the behaviour of $\xi$ must have a non-trivial kernel. Now the kernel of such an operator should be finite dimensional, and may certainly be trivial. The necessary condition that $\mathcal{D}$ have a non-trivial kernel highly constrains the possibility of the existence of Ricci solitons. 
Indeed long ago Bourguignon \cite{Bourguignon} showed that there are no Ricci solitons on a compact manifold without boundary for any choice of vector $\xi$.
In certain cases such as for asymptotically flat or Kaluza-Klein metrics, as we shall discuss later we may prove the kernel is again trivial and solitons cannot exist for the DeTurck choice of $\xi$ we are interested in. One arrives at the surprising conclusion that despite the fact that one is solving the Harmonic Einstein equation, which naively is quite different from the Einstein equation, in certain situations of interest the only solutions are in fact Ricci flat ones with the gauge condition $\xi = 0$ imposed.

Even if solitons do exist this is not a problem in principle.
Since the Harmonic Einstein equation is elliptic, for well posed boundary conditions on the metric we expect that solutions are locally unique. Hence a solution cannot be continuously deformed into another solution without suitably adjusting the boundary conditions. If there exists a Ricci flat solution, then any Ricci soliton solutions cannot be arbitrarily `nearby' to it. Hence numerically it should always be possible to distinguish the solution of interest from the solitons.
An obvious test is to compute the vector field $\xi$ and see if it is zero. In particular let us define the scalar $\phi \equiv \xi^\mu \xi_\mu$ so that it gives the norm of $\xi$. For a Riemannian manifold vanishing $\xi$ is necessary for the function $\phi$ to vanish. Hence we may check the magnitude of $\phi$ computed for our solution, and if it is anywhere non-zero then the solution is a soliton and we should try to find another one. 
An important question in practice is how many soliton solutions there are. If one were trying to find a single Ricci flat solution in a vast forest of solitons then such an approach may be impractical. If it is a single Ricci flat solution in a small spinney it is practical. Obviously if no solitons exist then it is ideal.

\subsection{Asymptotically flat or Kaluza-Klein solutions}

Treating our static black hole problem as a Riemannian boundary value problem we must impose boundary conditions asymptotically. Our discussion now follows that developed in \cite{PFLuciettiTW} with Figueras and Lucietti.
We impose that the Riemannian geometry has an asymptotic region such that the metric approaches the direct product $S_{\beta}^1 \times \mathcal{M}$, where $S^1_\beta$ is the Euclidean time circle with length $\beta$ (corresponding to temperature $T = 1/\beta$) and $\mathcal{M}$ is a Ricci flat manifold, the obvious choices being Euclidean space $\mathbb{R}^{D-1}$ or $\mathbb{R}^{D-2} \times S^1_L$ the product of Euclidean space with a circle of length $L$. The former choice is appropriate for the Euclidean continuation of asymptotically flat Lorentzian spacetime, and the latter would be for asymptotically Kaluza-Klein spacetime. 

Let us consider the case where we require the black hole to be asymptotically flat. We require the manifold to have an asymptotic region where for some large $R$ the metric behaves as,
\begin{eqnarray}
ds^2 = g_{\mu\nu} dx^\mu dx^\nu &=& d\tau^2 + \delta_{ij} dx^i dx^j + O(r^{-p}) \nonumber \\
\partial_i g_{\mu\nu} &=& O(r^{-p-1}) \, , \quad 
\partial_i \partial_j g_{\mu\nu} = O(r^{-p-2})
\end{eqnarray}
for all $r > R$ and for some positive $p$. Here $x^i$ are the usual Euclidean coordinates with $r=\sqrt{\delta_{ij} x^{i} x^{j}}$. For a Ricci flat solution we expect $p = D-3$. We require that the reference metric $\bar{g}$ which defines the vector $\xi$ as in \eqref{eq:xi} is also asymptotically flat, so that in the same part of the manifold this reference metric also behaves as above.

In practice it is convenient to compactify the radial coordinate in this asymptotic region. Taking $\rho = 1 / r$, we write the metric as,
\begin{eqnarray}
ds^2 &=&  N^2 d\tau^2 + \frac{\alpha^2}{\rho^4} d\rho^2 + \frac{1}{\rho^2} h_{ab} ( d \theta^a + w^a d\rho )( d \theta^b + w^b d\rho )
\end{eqnarray}
and require that, $N = 1$, $\alpha = 1$, $\omega_a = 0$ and $h_{ab} =  \Omega_{ab}$ at $\rho = 0$, where $\Omega_{ab}$ is the unit $(D-2)$-sphere metric. These appear as Dirichlet boundary conditions at $\rho = 0$ for the components of $g$, although we should recall that $\rho = 0$ is really a regular singular point of the PDEs, since this is an asymptotic region rather than a boundary at finite distance.

One may compute straightforwardly that taking an asymptotically flat reference metric, then,
\begin{equation}
\xi^{\tau}= O(r^{-p-1}) \qquad \xi^{i}= O(r^{-p-1}) \, .
\end{equation}
We see that the norm $\phi = \xi^\mu \xi_\mu \sim O(r^{- 2 p-2})$ so that the vector $\xi$ goes to zero length asymptotically for any positive value of $p$. Thus these asymptotic boundary conditions are consistent with the linear elliptic problem $\mathcal{D} \cdot \chi = 0$ discussed in the previous section being well posed and having trivial solution.
The asymptotically Kaluza-Klein case proceeds in exactly the same manner, where one now requires that the metric asymptotes to,
\begin{eqnarray}
g &=& d\tau^2 + \delta_{ij} dx^i dx^j + d y^2 + O(r^{-p}) \nonumber \\
\partial_i g_{\mu\nu} &=& O(r^{-p-1})  \, , \quad 
\partial_i \partial_j g_{\mu\nu} = O(r^{-p-2})
\end{eqnarray}
for some positive $p$
where $y$ is the compact Kaluza-Klein circle, and has period $L$. One obtains the same behaviour, $\phi \sim O(r^{- 2 p-2})$, as in the asymptotically flat case.

\subsection{A maximum principle and the non-existence of solitons}

We have discussed above that a static asymptotically flat or Kaluza-Klein black hole may be thought of as a smooth Riemannian manifold with no boundary except for the asymptotic region. We have shown that the vector $\xi \rightarrow 0$ in this asymptotic region provided the reference metric shares the same asymptotics. We now turn to the important question raised earlier of whether we can control the existence of Ricci solitons with such boundary conditions. Following work with Figueras and Lucietti \cite{PFLuciettiTW} we will now show that solitons cannot exist as we claimed earlier. This implies that solving the Harmonic Einstein equation is equivalent to solving the Einstein equation together with imposing the generalised harmonic gauge condition $\xi = 0$.

Contracting the Bianchi identity \eqref{eq:Bianchi} with the vector and using the soliton equation  yields,
\begin{eqnarray}
\label{eq:phicondition}
\nabla^2 \phi + \xi^\mu \partial_\mu \phi = ( \nabla_\mu \xi_\nu ) ( \nabla^\mu \xi^\nu ) \ge 0
\end{eqnarray}
where again $\phi = \xi^\mu \xi_\mu$ is the norm of the vector field and we have used the fact that for a Riemannian geometry the right-hand side is non-negative everywhere. Suppose now we have a solution to the Harmonic Einstein equation. Consider a function $f$ obeying the linear elliptic equation,
\begin{eqnarray}
\label{eq:maxeq}
\nabla^2 f + \xi^\mu \partial_\mu f  \ge 0 \, .
\end{eqnarray}
Now this equation enjoys a maximum principle, which states that if $f$ is non-constant, it must attain its maximum value on the boundary of the manifold. 
\footnote{
See for example \cite{Protter}, or for a statement of results specifically on Riemannian manifolds see \cite{Aubin}.
} 
Furthermore, if a maximum exists at the boundary the outer normal derivative of $f$ at this maximum is strictly positive.

Since the function $\phi \ge 0$, and since a Ricci soliton must have $\phi \ne 0$, then a necessary condition for a solution to be a soliton is that $\phi$ is either constant and non-zero or it has a maximum somewhere. If $\phi$ is constant and non-zero, then this implies $\nabla_\mu \xi_\nu = 0$ and so the vector $\xi$ is covariantly constant, and in fact whilst the solution is a soliton it is also Ricci flat. Thus a challenge to Ricci flatness means that $\phi$ must be non-constant and hence have a maximum somewhere. Since replacing $f$ with $\phi$ in \eqref{eq:maxeq} we obtain the condition \eqref{eq:phicondition}, and for \eqref{eq:maxeq} we have a maximum principle, we see that if $\phi$ is non-constant then its maximum must be at the boundary of the manifold, with positive normal outer gradient, or in an asymptotic region.

In the case of no boundary, we recover the result that a solution must be Ricci flat. For black hole solutions which necessarily have some form of boundary, either at finite proper distance from the horizon or asymptotically at infinite proper distance, we see that the existence of a soliton is intimately tied to the precise form of the boundary conditions imposed there. 

Let us consider the asymptotically flat or Kaluza-Klein boundary conditions discussed in the previous section. Assuming we find a solution to the Harmonic Einstein equation compatible with those asymptotics we calculated that $\phi = | \xi |^2 = O(r^{-2 p-2})$ and hence for the very weak requirement that $p>0$, this implies $\phi \rightarrow 0$ asymptotically. Provided there are no other boundaries, a simple application of the maximum principle rules out the existence of Ricci solitons as we claimed earlier.\footnote{
Assume for contradiction that there is a Ricci soliton so that $\phi \ne 0$. Consider some $R' > R$, so that on the closed surface $r = R'$ in the asymptotic region we have $\phi \leq C/R'^{2p+2}$ for some constant $C$. Then everywhere in the interior of this surface the maximum principle states that $\phi \leq C/R'^{2p+2}$. Under the assumption that $\phi \ne 0$, we must be able to find a point $p$ where $\phi$ is non-zero, and denote its value $\phi_0$, so $\phi_0 = \phi(p)$. Choosing $R'$ large enough, we may always ensure that the point $p$ is in the interior of the portion of the manifold bounded by the surface $r = R'$ and also that $\phi_0 > C/R'^{2p+2}$. Since we know in this interior $\phi \leq C/R'^{2p+2}$ we are lead to a contradiction and must conclude that $\phi = 0$ everywhere as a consequence of the maximum principle. 
}

\subsection{Isometries and `fictitious' boundaries}
\label{sec:fictitious}

Let us now turn to a more practical issue, namely using isometries to reduce the effective dimension of the PDEs. A PDE problem in high dimensions rapidly becomes intractable. 
Thus if we have isometries in our geometry it is important to manifest them explicitly using adapted coordinates and hence benefit from the storage savings of not having represented the metric in the isometry directions. The example of the 5D Kaluza-Klein black holes discussed in Chapter 4 is a case in point. This 5D problem is reduced to 2 effective dimensions when one takes into account the static $U(1)$ and rotational $SO(3)$ isometries.

Suppose we wish to numerically represent a function that has spherical symmetry in Euclidean space $\mathbb{R}^m$. Using Cartesian coordinates, so $ds^2 = \delta_{ij} dx^i dx^j$, the origin of the symmetry $x^i = 0$ is not a special point and the function is simply smooth there. If we use coordinates adapted to the symmetry, spherical polar coordinates $ds^2 = dr^2 + r^2 d\Omega^2$, the function only depends on the radial coordinate $r$ giving the saving in storage. However, we find that the polar coordinate chart breaks down at the origin, and we must treat the point at the origin effectively as a boundary point - we term this a `fictitious boundary'. 
Of course we may deduce the required boundary conditions in the polar chart simply from the requirement that in the Cartesian coordinates the function is smooth, so that $f(x)$ is a $C^\infty$ function. Then using the fact that $r^2 = \delta_{ij} x^i x^j$ this implies a function that only has radial dependence must be a smooth function in $r^2$, i.e. $f = f(r^2)$ is $C^\infty$. 

There are two important fictitious boundaries that typically arise when considering static black holes. Firstly associated to the Euclidean static $U(1)$ isometry we have that the horizon is precisely the place where the time circle vanishes, and hence the isometry has a fixed action. In this case we may write the metric in polar coordinates as,
\begin{eqnarray}
\label{eq:Euchoriz}
ds^2 =  A dr^2 + r^2 B d\tau^2 + r C_a dr dx^a + h_{ab} dx^a dx^b 
\end{eqnarray}
where the component functions only depend on the radial and $x$ coordinates, and are independent of the time direction $\tau$ which we recall has period $\tau \sim \tau + 2 \pi / \kappa$.
Using the Cartesian coordinates $X$ and $Y$ mentioned above, so $X = r \cos{\kappa \tau}$ and  $Y = r \sin{\kappa \tau}$, then one can show the metric components in these coordinates are smooth functions of $X$ and $Y$ at the fixed point $r = 0$ provided that $A, B, C_a, h_{ab}$ are smooth functions of $r^2 = X^2 + Y^2$ and the coordinates $x$, and furthermore that $\kappa^2 A = B$ at $r = 0$ for some constant $\kappa > 0$. Here the constant $\kappa$ gives the surface gravity of the horizon with respect to $\partial/\partial \tau$.

Since there is not a real boundary at the horizon our maximum principle must ensure no maximum of $\phi$ can occur there. Treating the horizon in static adapted coordinates we see the horizon as a fictitious boundary. We may simply compute that $\xi^r = \partial_r \xi^a = 0$ at $r = 0$. Now $\xi^\tau = 0$ everywhere and hence $\partial_r \phi|_{r=0} = 0$. Recall the maximum principle implies a maximum at a boundary requires $\phi$ has positive outer normal gradient. We see since $\partial_r \phi = 0$ at $r=0$, the maximum principle does indeed rule out a maximum at this fictitious boundary.

Secondly one typically has rotational axisymmetry in the problem and an associated $SO(n)$ isometry group. The axis of this symmetry is then the set of points fixed under the isometry. In this case we write the polar metric as,
\begin{eqnarray}
ds^2 =  A dr^2 + r^2 B d\Omega^2 + r C_a dr dx^a + h_{ab} dx^a dx^b 
\end{eqnarray}
where $d\Omega^2$ is the line element on a unit $(n-1)$-sphere. Transforming to appropriate Cartesian coordinates
one finds a smooth metric again provided $A, B, C_a, h_{ab}$ are smooth functions of $r^2$ and $A = B$ at $r=0$. Note that there is no free constant as in the case of the vanishing $U(1)$ as we have already chosen the sphere $d\Omega^2$ to have unit radius. As above, no maximum can reside here as $\partial_r \phi = 0$ at $r=0$.

\subsection{Solving the Harmonic Einstein equation I: Ricci flow as local relaxation}

We shall now consider the canonical method to solve an elliptic system, and find that in fact for the Harmonic Einstein equation this can be thought of in the continuum as the famous Ricci flow.

Suppose we consider the Laplace equation, $\nabla^2 \psi = 0$, in some finite region $U$ of $D$-dimensional Euclidean space, and wish to solve this boundary value problem with some appropriate boundary conditions (e.g. Dirichlet). The simplest numerical approach is to represent the function $\psi$ using real space finite difference and use local relaxation to find a solution. 
Let us take the canonical coordinates $(x_1,x_2,\ldots,x_D)$ on $\mathbb{R}^D$, and consider the rectangular lattice of points with lattice spacing $\Delta$ at positions $( m_1 \Delta , m_2 \Delta , \ldots , m_D \Delta)$ for integers $m_1, m_2 \ldots, m_D$. Denote the set of points $\{ p_i \}$ in this lattice that lie in the interior of the domain $U$ as $L$, where $i = 1 , \ldots , N$ labels these points. Real space finite difference represents the function $\psi$ by storing the set of values $\{ \psi_{i} \}$ where $\psi_i = \psi( p_{i} )$. 

The simplest way to represent the Laplace equation is by second order finite difference. 
Consider a point $p_i \in L$. We denote the $2 D$ nearest neighbour points in this rectangular lattice as $p_{i->j}$, where $j = 1, 2, \ldots , 2 D$. Then we may approximate,
\begin{eqnarray}
\nabla^2 \psi |_{p_i}  \simeq  \frac{1}{\Delta^2} \left( - (2 D) \psi_i + \sum_{j=1}^{2D} \psi_{i->j} \right) \, .
\end{eqnarray}
By making $\Delta$ smaller, we approximate this Laplacian increasingly well, and we now have a finite representation of the continuum Laplace equation. Note that for points in $L$ that neighbour points in the boundary or exterior to $U$, then we consider the values of those neighbours to be fixed by the boundary conditions.

We then proceed by solving this finite problem. The classic method is that due to Jacobi, a local iterative procedure also known as relaxation. 
Let us imagine a sequence of guesses to the solution, $ \{ \psi^{(A)}_{i} \} $ for integer $A=0,1,2,\ldots $.
The Jacobi method states that given a guess $\{ \psi^{(A)}_{i} \}$ we can improve it by computing a new guess $\{ \psi^{(A+1)}_{i} \}$  as
\begin{eqnarray}
\label{eq:Jacobi}
\psi^{(A+1)}_i = \frac{1}{2 D}\left( \sum_{j=1}^{2D} \psi^{(A)}_{i->j} \right) 
\end{eqnarray}
The idea is that one takes an initial guess $\{ \psi^{(0)}_i \}$ satisfying the boundary conditions, and then iterates Jacobi's improvement method. For the simple Laplace equation one will reach a fixed point of this iteration which is a solution of the finite Laplace equation above.

We may rearrange the equation above as,
\begin{eqnarray}
\frac{2 D }{\Delta^2} \left( \psi^{(A+1)}_i - \psi^{(A)}_i \right) = \frac{1}{\Delta^2} \left( \sum_{j=1}^{2 D} \psi^{(A)}_{i->j} - (2 D) \psi^{(A)}_i  \right) 
\end{eqnarray}
but now this may be viewed as the finite differencing of the diffusion equation, where now $\psi$ is a function of $x$ and a flow time $\lambda$, and,
\begin{eqnarray}
\pd{\psi(\lambda,x)}{\lambda} = \nabla^2 \psi(\lambda,x) 
\end{eqnarray}
where we discretize time similarly to space, so that $\psi^{(A)}_i = \psi( A \, \delta ) |_{p_i}$ where $\delta = \frac{\Delta^2}{2 D}$ gives the discretization spacing in flow time. The right-hand side is approximated as above, and the left-hand side is approximated using forward Euler differencing,
\begin{eqnarray}
\left. \pd{\psi}{\lambda} \right|_{\lambda = A \delta} \simeq \frac{1}{\delta} \left( \psi^{(A+1)}_i - \psi^{(A)}_i \right)  \, .
\end{eqnarray}
Thus we see that the simplest local relaxation method used to solve the Laplace equation is in fact simply diffusion on spatial and temporal scales must larger than the lattice scale $\Delta$. Instead of thinking about Jacobi local relaxation and finite difference of the elliptic problem, we can describe this process more formally as the parabolic continuum diffusion problem. Then we take an initial guess, and act on it with diffusion until we reach a fixed point of the diffusion flow, and this will solve the original elliptic problem.

We may use the Jacobi method for the more complicated elliptic Harmonic Einstein equation. Instead of a single function $\psi$, in a chart there are all the components of the metric tensor $g_{\mu\nu}$ to solve for. We may represent the metric in a chart by again taking the same lattice $L$ in $\mathbb{R}^D$, and we store the set of values $(g_{\mu\nu})_i \equiv g_{\mu\nu}( p_i )$. Recall that from \eqref{eq:symbol} the Harmonic Einstein equation has second derivative structure,
\begin{equation}
R^H_{\mu\nu} = - \frac{1}{2} g^{\alpha\beta} \partial_\alpha \partial_\beta g_{\mu\nu} + T_{\mu\nu}
\end{equation}
where $T_{\mu\nu}$ represents the terms with lower numbers of derivatives. Jacobi's method is then the discretisation of the continuum equation,
\begin{eqnarray}
\label{eq:ricciflow2}
\pd{g_{\mu\nu}(\lambda)}{\lambda} = - 2 R^H_{\mu\nu}
\end{eqnarray}
where the right-hand side is approximated using second order finite difference, and the left-hand side is differenced using the forward Euler method. We may think of Jacobi as approximating the Harmonic Einstein equation as a set of Poisson equations, where the sources in the Poisson equations are the lower derivative terms $T_{\mu\nu}$. It attempts to
 solve the Poisson equations at each point in isolation by considering the neighbours and sources as fixed.

In fact there are many variants of the classical Jacobi method of relaxation for elliptic systems. For example Gauss-Seidel speeds up Jacobi by updating the values of the points in situ, rather than storing a whole new approximation $\psi^{(A+1)}$ given a previous one $\psi^{(A)}$. This is again another discretisation of the continuum diffusion problem although with an exotic time derivative differencing. Over and under relaxation schemes simply adjust the diffusion constant. Multi-grid methods work by concurrently solving the equation on multiple lattices, cascading information from low to high resolution and back, in order to allow information to propagate quickly over long distances. These methods give great enhancement of speed, but on scales larger than the lowest resolution lattice, again act as continuum diffusion.

Thus whichever local relaxation method one chooses, one may consider the problem to be continuum diffusion on large enough scales, and for the Harmonic Einstein equation, the canonical diffusion is the flow in time $\lambda$ given in equation \eqref{eq:ricciflow2}, which explicitly is,
\begin{eqnarray}
\label{eq:RDeTflow}
\pd{g_{\mu\nu}(\lambda)}{\lambda} = - 2 R_{\mu\nu} + 2 \nabla_{(\mu} \xi_{\nu)} \, .
\end{eqnarray}
In fact this is precisely the Ricci-DeTurck flow, where the second term is an infinitesimal diffeomorphism generated by the vector field $\xi$, so that the flow is diffeomorphic 
\footnote{
In the presence of boundaries we should ensure that the normal component of $\xi$ at a boundary vanishes in order that these two flows are diffeomorphic. This ensures that the diffeomorphisms generated by $\xi$ along the flow act to preserve the boundary points.
}
to the Ricci flow
\begin{eqnarray}
\label{eq:Ricciflow}
\pd{g_{\mu\nu}(\lambda)}{\lambda} = - 2 R_{\mu\nu}
\end{eqnarray}
introduced by Hamilton as a tool in geometric analysis, and which has gained fame for its role in proving the Poincar\'e conjecture. For an introduction to Ricci flow see for example \cite{Topping}.
DeTurck proved that Ricci flow was a well posed parabolic flow by realising that one can add the diffeomorphism term and with the choice in \eqref{eq:DeTurck} explicitly render the flow equation parabolic.

Again on scales larger than the lattice resolution used in the local relaxation, we may view the methods discussed above formally as the Ricci-DeTurck flow. We may regard this parabolic flow as a continuum algorithm to solve the Harmonic Einstein equation. We give some initial guess for the parabolic flow. A fixed point of the flow is a solution of the Harmonic Einstein problem. Hence, by simulating the flow for sufficient flow time, we might hope to approach a fixed point as closely as we require. 

One beautiful consequence of this is that whilst we required some choice of reference metric to define the vector field $\xi$ and render the Harmonic Einstein equation elliptic, and hence well posed as a boundary value problem, in fact the Ricci-DeTurck flow is diffeomorphic to Ricci flow which makes no reference to $\xi$. Thus given some initial guess metric, whilst different choices of reference metric will change the path taken in the space of metrics by the Ricci-DeTurck flow, the path taken in the space of geometries (i.e. metrics modulo diffeomorphisms) is always the same. 

Provided one chooses the reference metric to share the same isometries as the metric, for example the static symmetry, then the Harmonic Einstein tensor is also symmetric under these isometries. This implies that the Ricci-DeTurck flow preserves the isometries of the metric. 

Ricci flow has very nice properties geometrically. Essentially it is diffusion for geometry, and locally tries to smooth out curvature. Suppose we have a Ricci flat solution, $g_{\mu\nu}$, and wish to consider the Ricci flow of a perturbation to this, $h_{\mu\nu}$. Then from \eqref{eq:linRicci} we see,
\begin{equation}
\pd{h_{\mu\nu}}{\lambda} = -2 \Delta_L h_{\mu\nu} -2 \nabla_{(\mu}v_{\nu)} \,,
\end{equation}
where $\Delta_L$ is the Lichnerowicz operator on the background $g$,  
and the last term is simply an infinitesimal diffeomorphism generated by $v$. For Euclidean space, this flow is then diffeomorphic to a flow where each component of the metric simply diffuses 
as, $\pd{h_{\mu\nu}}{\lambda} = \delta^{\alpha\beta} \partial_\alpha \partial_\beta h_{\mu\nu}$. Hence we see that Euclidean space is stable to linear perturbations. However beyond this diffusive behaviour of linear perturbations  it has very interesting non-linear properties. For example, it wishes to collapse regions of positive curvature as can be seen by the Ricci flow of a round sphere, whose radius shrinks linearly in flow time, reaching a zero size in finite time. Of relevance for us, it has been rigourously proven that Ricci flow exists and preserves asymptotic flatness for short times \cite{ToddOlnyik}.

An important property of the Ricci flow is that we see from above a Ricci flat solution is a stable fixed point only if the operator $\Delta_L$ is positive. We may assume there are no zero modes, as these should be fixed appropriately by boundary conditions, which should ensure a locally unique solution. However $\Delta_L$ may not be positive in general, so that there exist one or more eigenfunctions with negative eigenvalue. At late times such eigenmodes exponentially grow in time so that the perturbation will flow one away from the fixed point in these directions.

A fundamental property of static vacuum black holes is that the positivity of $\Delta_L$ for their Euclidean continuation is related to their thermodynamic behaviour, since the Euclidean action is simply related to their free energy. In the asymptotically flat case this has been made precise in \cite{Dias:2010eu}. In many cases static black holes of interest (for example, all those in Kaluza-Klein theory discussed in Chapter 4) possess negative modes of $\Delta_L$, the canonical example being that of the asymptotically flat Euclidean Schwarzschild solution which has the single negative mode discovered by Gross, Perry and Yaffe \cite{GPY}. 

We have seen that standard local relaxation methods can be viewed as Ricci-DeTurck flow on large scales, which is diffeomorphic to Ricci flow. Thus for all these methods, one cannot simply start with an initial guess and flow to the black hole fixed point solution if it possesses  negative modes. Starting with an initial guess close to the solution one will flow towards the fixed point in nearly all directions in the space of perturbations of the fixed point, but will generically then veer off along the direction(s) tangent to the negative mode(s). 

Since many solutions of interest have negative modes one might imagine that local relaxation hopelessly fails to provide an algorithm to find solutions of the elliptic Harmonic Einstein problem. This conclusion however is too quick, and in principle Ricci flow and hence local relaxation may still be used but the method must be modified slightly. Suppose the fixed point we are interested in has a single negative mode of $\Delta_L$ and up to diffeomorphisms has no zero modes so that the fixed point is locally unique. Examples of this include localised black holes in Kaluza-Klein theory discussed in Chapter 4.
Let us call the fixed point $g_0$. Then locally about $g_0$ the space of geometries (meaning metrics up to diffeomorphisms) is infinite dimensional.
There are two special flows that emanate from $g_0$ along the negative mode direction. Let us denote the negative mode perturbation $h$ at $g_0$. 
Then these two flows are $g_{\pm}(\lambda)$ such that $g_{\pm} \simeq g_0 \pm e^{+ 2 v^2 \lambda} h$ as $\lambda \to - \infty$, where $\Delta_L h = - v^2 h$ so that $v^2$ is the magnitude of the negative eigenvalue.
The perturbation $h$ is a tangent vector to the space of geometries at $g_0$. A basis for the tangent space at $g_0$ is given by $h$ together with the positive eigenmodes of $\Delta_L$. The positive eigenmodes are tangent to a codimension one surface,  $\Sigma$, which contains $g_0$ and is closed under Ricci flow. Starting from any point in $\Sigma$ near to $g_0$ one remains within this surface under the action of the flow, and will flow to the fixed point reaching it asymptotically.

The problem is then to generate an initial guess contained in $\Sigma$ since this will flow to the fixed point. 
Consider a one parameter family of geometries $g(\alpha)$, where $\alpha$ is the parameter. It is generic that the curve $g(\alpha)$ will intersect $\Sigma$. Suppose this occurs at $\alpha = \alpha_\star$. Then for $\alpha > \alpha_\star$ but close to $\alpha_\star$ one will initially flow towards $g_0$ and then be carried away in the direction of the negative mode. In the limit that $\alpha \to \alpha_{\star}$, the flow will at late times follow that of either $g_+$ or $g_-$. Let us assume it follows $g_+$ for $\alpha > \alpha_\star$. Then conversely for $\alpha < \alpha_\star$ one will approach $g_0$ and then deviate away in the opposite sense, flowing away close to the flow $g_-$.

Thus we see that for $\alpha$ close to $\alpha_\star$ there is a critical behaviour associated to the unstable fixed point $g_0$. 
By scanning the values of $\alpha$ one can hope to see this critical behaviour, and if one can identify whether one has flowed in the $g_+$ or $g_-$ direction, one can simply automate a tuning of $\alpha$ to get as close to $\alpha_\star$ as required. Then one has flows that get very close to $g_0$ for a long period of flow time before finally succumbing to the negative mode and flowing away.  In principle one can get as close to the fixed point as desired. 

For a Schwarzschild black hole the flows $g_{\pm}$ generated by the negative mode either expand (say $g_+$) or contract ($g_-$) the horizon. The flow $g_-$ has been shown to continue to shrink the horizon to a finite time singularity, whilst under the flow $g_+$ the horizon grows without stopping \cite{HeadrickTW}. Thus given a flow it is very simple to see which side of $\Sigma$ that flow is on, and hence tune the parameter $\alpha$ to reach $\alpha_{\star}$. Similar behaviour is seen for the Kaluza-Klein localised black holes, where one flow pinches the horizon to zero size and a finite time singularity, and the other expands it until its poles touch and again a singularity is reached \cite{HeadrickKitchenTW}.

In principle this method may be extended to a case with $N$ negative modes. Then an $N$ parameter family of initial data must be tuned in order to reach the fixed point. In the codimension one case, so that $\Sigma$ partitions the space of geometries locally about $g_0$, it is easy to see which `side' of $\Sigma$ one starts on. This is not true in the higher codimension case, and one must simply search the space of parameters until one locates the critical point $\alpha_{\star}$.

Let us summarise this discussion. Local relaxation is the simplest method to solve elliptic PDEs, and can be applied to the Harmonic Einstein equation. On large scales we may think of relaxation from a continuum perspective as Ricci-DeTurck flow, which is diffeomorphic to Ricci flow. Hence this approach has the beautiful geometric property that the trajectory taken by the flow is \emph{independent} of the reference metric and hence the gauge fixing. 
However many black holes are unstable fixed points Ricci flow, and for these relaxation or Ricci flow may still be used to find these solutions, but 
for a solution with $N$ negative modes of its Lichnerowicz operator, one must find a suitable $N$ parameter set of initial data, and tune these $N$ parameters in order to flow or relax to a solution. We emphasize that there do exist interesting solutions with Killing horizons which are stable under Ricci flow, such as those in AdS/CFT where the boundary metric is a black hole \cite{PFLuciettiTW}.

\subsection{Solving the Harmonic Einstein equation II: Newton's method}

We have seen that whilst the simplest relaxation methods to solve elliptic systems have an elegant geometric behaviour on large scales, it is difficult to find many black holes of interest which are unstable fixed points of these methods due to having Euclidean negative modes. With one such negative mode these methods are still practical. For more they become increasingly hard to use.

Fortunately there is a second standard technique to solve these elliptic systems, namely Newton's method (also known as the Newton-Raphson method). As we shall see this approach is considerably more complicated to implement, and lacks the geometric elegance of relaxation, so that the behaviour of Newton's method will explicitly depend on the choice of reference metric. However, the advantage of Newton's method is that it is insensitive to the stability of the fixed point. In fact the basin of attraction of Newton's method can be rather small in practice, and thus a combination of Ricci flow or relaxation to get close to the fixed point, followed by Newton's method to hone in on it can be the best strategy.

Unlike relaxation, Newton's method is inherently non-local. Let us again imagine discretising our system using finite difference as above. At each lattice point in a chart we will have the various components of the metric. Globally there will be a finite set of numbers $\{ g_M \}$ that will give the finite difference approximation to the metric $g_{\mu\nu}(x)$, where the index $M$ includes both the lattice point in a given chart and the component of the tensor. Likewise we may represent the Harmonic Einstein tensor in the interior of the manifold with the same index structure. Then the Harmonic Einstein equation is given as $R^H_M(g)= 0$, which can be thought of as a finite set of coupled non-linear equations in the variables $g_M$. The canonical way to solve such a system is by the multidimensional generalisation of Newton's method.

If we perturb the metric $g$ as $g + \epsilon \, \delta g$, the Harmonic Einstein tensor goes as,
\begin{eqnarray}
R^H_M( g + \epsilon \, \delta g) = R^H_M (g) + \epsilon\, \mathcal{O}(g)_{M}^{~N} \delta g_{N} + O(\epsilon^2) \, ,
\end{eqnarray}
where the matrix $\mathcal{O}(g)_{M}^{~N}$ is the linearisation of $R^H_M$. Begin with an initial guess $g^{(0)}_M$. Then 
Newton's method iteratively improves a trial metric $g^{(A)}_M$ as,
\footnote{It is sometimes useful to take `smaller' steps, with $g^{(A+1)}  = g^{(A)} - \epsilon \,  \mathcal{O}(g^{(A)})^{-1} \cdot R^H(g^{(A)})$ for some $\epsilon$ with $0< \epsilon < 1$, particularly in the first iterations if the initial guess is not very close to the solution.}
\begin{eqnarray}
g^{(A+1)}_M  = g^{(A)}_M - (\mathcal{O}(g^{(A)})^{-1})_M^{~N} R^H_{N}(g^{(A)}) \, .
\end{eqnarray}
As with the one dimensional Newton method this moves along the tangent of the equations to find a solution. Near a solution it will very quickly converge to that solution. However, the basin of attraction may be rather small in practice, and outside of this iterations of Newton's method will usually diverge and give singularities.

As for the Ricci flow method, provided the reference metric is chosen to have the same isometries as the metric, then the Harmonic Einstein tensor will be symmetric under these and the Newton method will act to preserve these isometries.

This method has the important advantage over the Ricci-DeTurck flow method that it is not sensitive to negative modes of the Lichnerowicz operator. However it does assume that the linear problem $\mathcal{O} \cdot V = R^H$ can be solved for the vector $V$. In practice robust methods exist to solve such (finite dimensional) linear systems, such as biconjugate gradient, which are insensitive to the spectrum of $\mathcal{O}$, provided there are no zero modes which we assume for well posed boundary data. Thus a single initial guess will suffice, rather than having to tune a family of initial guesses.

We see that the implementation of Newton's method is considerably more complicated than that of the relaxation/Ricci flow methods. Another important disadvantage of the Newton method over the Ricci-DeTurck flow is that it is not geometric in the sense that the path taken by the algorithm in the space of geometries will depend explicitly on the choice of reference metric. This implies that the basin of attraction of a solution, which in practice may be rather small, will also depend on this choice of reference metric. Sometimes it is actually convenient to use a combination of the Ricci flow method together with the Newton method. The Ricci flow method is rather robust and can quickly get one reasonably close to a fixed point. It is tuning the flows very close to the fixed point that becomes difficult and time consuming. However, once reasonably close, one can simply use the Newton method to quickly find the precise fixed point.

\subsection{An illustrative example}
\label{sec:ex}

In order to illustrate many of the points discussed above we will now give a very simple example, where we already know the answer. We will consider finding the 4D Schwarzschild solution using the various techniques above. We will assume spherical symmetry, and so the problem really is one involving only ODEs but we shall treat it in an identical manner to the much more complicated PDE problems we are really interested in. This example is simple enough that we can be very explicit about the implementation which we detail below. 
\footnote{
We will make available a very simple {\it Mathematica} notebook that implements the relaxation / Ricci flow and the Newton algorithms in this toy example. Hopefully this provides an entry point for those interested in thinking about the more complicated problems of interest. This will be found at;  { \tt http://www3.imperial.ac.uk/people/t.wiseman }
}

We note that our earlier maximum principle argument states that in this asymptotically flat case no soliton solutions should exist and hence any solution to the Harmonic Einstein equation must be Ricci flat.
We will cover the manifold with one chart, and use the fact that the solution is static and spherically symmetric, adapting coordinates to these symmetries. We will choose a radial coordinate $r$, and choose the horizon to be located at $r = 0$, and infinity to be at $r = 1$. We continue to Euclidean time, and write the smooth Riemannian metric as,
\begin{eqnarray}
\label{eq:exmet}
ds^2 = r^2 A d\tau^2 + 4 f^2 B dr^2 + f C d\Omega^2 \; , \quad f = \frac{1}{(1-r^2)^2}
\end{eqnarray}
where $d \Omega^2 = d\theta^2 + \sin^2 \theta d\phi^2$. Then
 $A,B,C$ are functions of $r$, and the metric describes the general static spherically symmetric metric.
Our choice of factors above means that for the topology of manifold we wish to describe, $A , B , C> 0$ on the domain $r \in [ 0 ,1 ]$. 
At $r = 0$ we have a fictitious boundary as we have adapted coordinates to the static symmetry. 
From our previous discussion smoothness of the full Riemannian manifold implies that at $r = 0$ then $\kappa^2 = A / 4 B$ for surface gravity $\kappa$, and $A,B,C$ are smooth functions of $r^2$. At infinity, $r = 1$, we fix $A = B = C = 1$.

The Harmonic Einstein tensor is determined by the components $R^H_{\tau\tau}$, $R^H_{rr} $ and $R^H_{\theta \theta} $.
We may discretize this system using finite difference by choosing $(1+N)$ lattice of points at locations $r_i = i \Delta$ for $i = 0, 1, \ldots  N$ with $\Delta = 1/N$, so that $A_i = A(r_i)$, and likewise for $B$ and $C$. Our boundary conditions imply that $A_{N} = B_N = C_N = 1$. At the horizon we require smoothness in $r^2$ which implies that we may deduce the boundary values $A_0, B_0, C_0$ in terms of the interior points. For small $r$ the behaviour goes as constant plus quadratic in $r$, which gives $A_0 = \left( 4 A_1 - A_2 \right) / 3$, and likewise for $B_0$ and $C_0$. We choose to also impose the regularity condition $A = B$ directly by taking $A_0 = B_0$. This in turn determines $B_1$ from the smoothness. In full we have,
\begin{eqnarray}
&& A_0 = \frac{1}{3} \left( 4 A_1 - A_2 \right) \, , \quad B_0 =  \frac{1}{3} \left( 4 A_1 - A_2 \right) \, , \quad C_0 =  \frac{1}{3} \left( 4 C_1 - C_2 \right)  \, , \nonumber\\
&& B_1 = A_1 + \frac{1}{4} (B_2 - A_2)  \, , \quad   A_{N} = B_N = C_N = 1
\end{eqnarray}
and the vector $g_M \equiv \{ A_1 \ldots A_{N-1} , B_2 \ldots B_{N-1} , C_1 \ldots C_{N-1} \}$ describes the metric subject to the above conditions. We finite difference derivative terms using simple second order differencing so that, 
\begin{eqnarray}
\partial_r X_i &=& \frac{1}{2\Delta} \left( X_{i+1} - X_{i-1} \right) \nonumber \\
\partial^2_r X_i &=& \frac{1}{\Delta^2} \left( X_{i+1} + X_{i-1} - 2 X_{i} \right) 
\end{eqnarray}
and then we may evaluate ${R^H_{\mu\nu}}_i \equiv R^H_{\mu\nu}( r_i )$.

Consider $g^{(0)}_M$ to be an initial guess, and $g^{(A)}_M$ to be subsequent iterations of improvement for $A = 1, 2, \ldots$. Then the Jacobi method, or equivalently the Ricci-DeTurck flow discretised in time using forward Euler differencing, gives,
\begin{eqnarray}
 r_i^2 A^{(A+1)}_i &=& r_i^2 A^{(A)}_i - 2 \,\delta \,{R^H_{\tau\tau} ( g^{(A)} ) }_i \; , \quad i = 1,\ldots,{N-1} 
\end{eqnarray}
and similarly for $B$ and $C$ except that for $B$ we have $i = 2, \ldots, {N-1}$. The continuum Ricci flow time $\lambda$ for $g^{(A)}$ is then given as $\lambda = A \,\delta$ and $\delta = \Delta^2/2$.
For the Newton method one creates the $(3 N - 4)$ vector of equations,
\begin{eqnarray}
R^H_M \equiv \left\{ {R^H_{\tau\tau}}_1 \ldots {R^H_{\tau\tau}}_{N-1} , {R^H_{rr}}_2 \ldots {R^H_{rr}}_{N-1} , {R^H_{\theta \theta}}_1 \ldots {R^H_{\theta\theta}}_{N-1} \right\}
\end{eqnarray}
which is a function of the $(3 N - 4)$ component vector $g_M$. Then the linearisation $\mathcal{O}_M^{~N} \equiv \partial R^H_M / \partial g_N$ is a square matrix, which can be inverted to solve the linear system required for the Newton method. 

Without loss of generality we choose $\kappa = 1/2$ by global scaling. 
Now we note at this point that $A(r) = B(r) = C(r) = 1$ is the Schwarzschild solution for $\kappa = 1/2$. Whilst this is a toy example, it almost seems too trivial to find the solution in these coordinates, and hence to challenge ourselves we will choose the background metric such that for the Schwarzschild solution the metric functions are not simply constant!  

We know that Schwarzschild is unstable to Ricci flow with one negative mode. Thus let us choose a one parameter family of initial metrics,
\begin{eqnarray}
A = 1 - \alpha \left( 1 - r^2 \right)^2  \, , \quad B  =  1  - \alpha \left( 1 - r^2 \right)^2  \, , \quad C =  \frac{1}{2} \left( 1 + r^2 \right) 
\end{eqnarray}
parameterised by the constant $\alpha$ that satisfy our boundary conditions for $\kappa = 1/2$. We choose the metric at zero flow time to be given by $A$, $B$ and $C$ above, and also take the fixed reference metric to be the same. We note that since $C$ above is not constant then the actual Schwarzschild solution has non constant metric functions in the generalized harmonic coordinates the reference metric imposes.

\begin{figure}[h]
\makebox[\textwidth]{
\includegraphics[width=12cm]{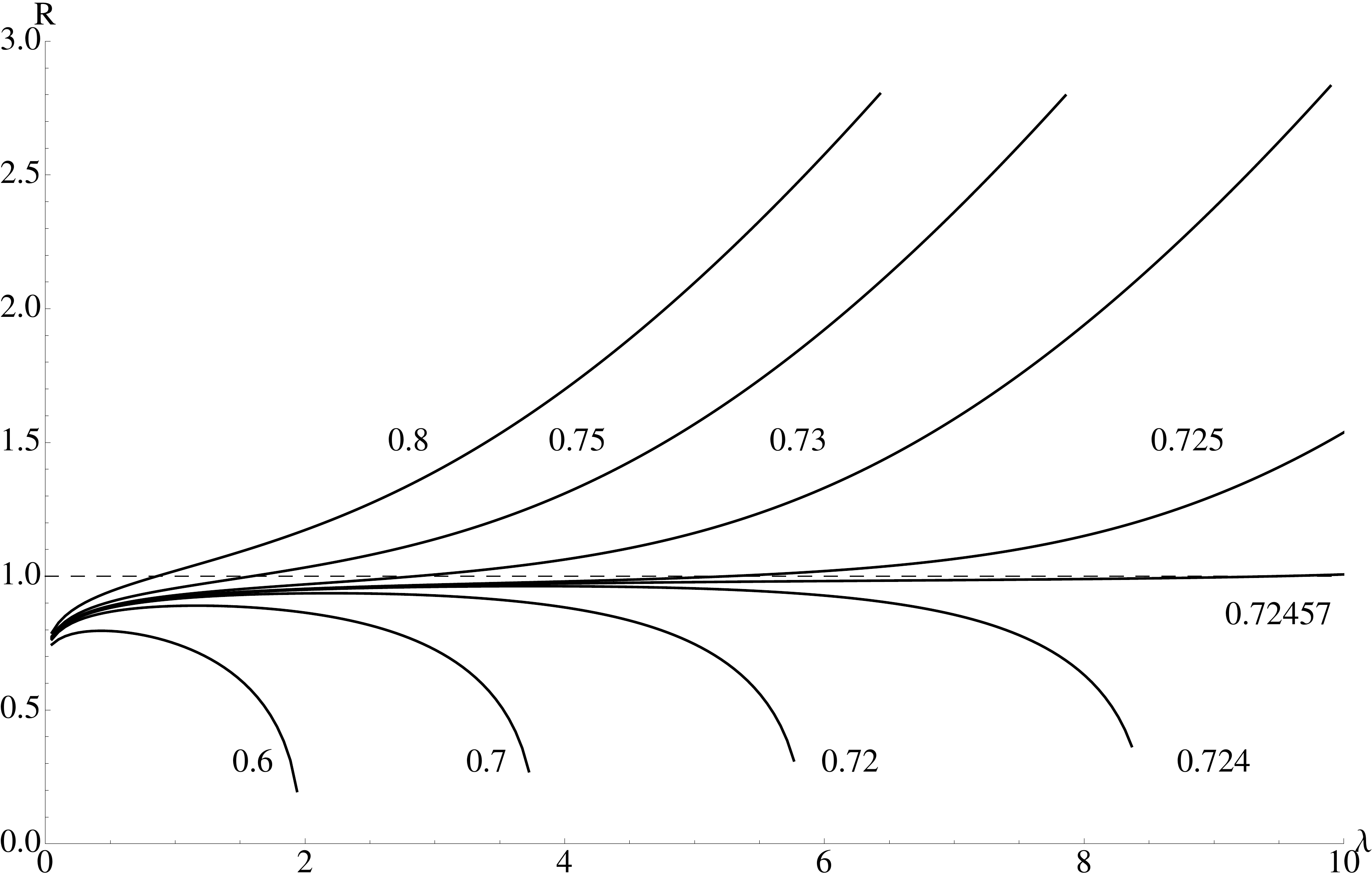}
}
\caption{
Figure depicting use of relaxation or the Ricci flow method to find the Schwarzschild solution. The plot shows the evolution of the radius of the horizon, $R$, as a function of flow time $\lambda$ for flows with a variety of values of the parameter $\alpha$. The values of $\alpha$ are labelled next to the corresponding curve.
After initial transient behaviour we see that for $\alpha < \alpha_\star \simeq 0.72$ the horizon shrinks (in fact to a singularity) at finite flow time. Conversely for $\alpha > \alpha_\star$ the horizon grows indefinitely. Tuning close to $\alpha_\star$ one may approach the Schwarzschild solution as accurately as desired. We see the quantity plotted here does indeed tend to one (the Schwarzschild value for the chosen surface gravity) for a finely tuned flow. (The data presented was computed using a very modest $N = 100$; we note the value of $\alpha$ required to fine tune the data will depend on this discretization, giving $\alpha_\star$ only in the continuum limit).
}
\label{fig:flow}
\end{figure}

Under relaxation/Ricci flow we might hope there is a critical value of $\alpha$, say $\alpha_\star$ where we may approach the unstable Schwarzschild fixed point. Indeed one finds there is, with the critical value $\alpha_\star \simeq 0.72$. In figure \ref{fig:flow} we plot the size of the sphere at the horizon $R \equiv \sqrt{C} |_{r = 0}$ against flow time $\lambda$ for a number of flows with various $\alpha$ approaching $\alpha_\star$ from above and below. We see that two very distinct behaviours are found for $\alpha > \alpha_{\star}$ or $\alpha < \alpha_{\star}$, that are straightforward to identify, and hence tune to the fixed point. Needless to say, the fixed point itself is indeed Schwarzschild, and we see that for $\alpha \simeq \alpha_{\star}$ the horizon does indeed tend to unit radius at late times as it should for Schwarzschild with $\kappa = 1/2$. 

The Newton method efficiently hones in on the Schwarzschild solution provided one is in the basin of attraction of the fixed point. In fact for the choice of initial metric and reference metric given above, a rather wide range of values of $\alpha$ all lie in the basin of attraction. For example, simply taking the initial guess with $\alpha = 0$ will quickly find the Schwarzschild solution after a handful of Newton iterations.

\section{Stationary vacuum solutions}

Generally we wish to be able to tackle stationary solutions and in this section we address how to extend the static methods discussed above to this case.
We note that the classic 4D uniqueness theorems relied on formulating the stationary axisymmetric problem as an elliptic system \cite{Carter}. Our task here is to formulate the general stationary vacuum problem as an elliptic system using the covariant Harmonic Einstein equation approach, and then ensure that the Ricci flow and Newton method algorithms may still be applied.
 This will require us to tackle the problem from a manifestly Lorentzian point of view. This section is based on the recent work \cite{newpaper} together with Adam and Kitchen.

\subsection{Static solutions from a Lorentzian perspective}

Instead of immediately considering stationary spacetimes, it is instructive to first consider static spacetimes from a Lorentzian perspective. The  Harmonic Einstein equation is not elliptic for a general Lorentzian manifold but rather it is hyperbolic, and without ellipticity one would not expect to be able to impose the various boundary conditions that physically we require in a well posed manner. However, consider a chart away from any horizon which manifests the static symmetry, 
\begin{eqnarray}
\label{eq:staticLor}
ds^2 = g_{\mu\nu} dx^\mu dx^\nu =  - N(x)^2 dt^2 + h_{ij}(x) dx^i dx^j
\end{eqnarray}
so that $N^2 > 0$. With the choice  that our reference metric is also static with respect to $\partial / \partial t$, so that,
\begin{eqnarray}
\bar{ds}^2 = \bar{g}_{\mu\nu} dx^\mu dx^\nu =  - \bar{N}(x)^2 dt^2 + \bar{h}_{ij}(x) dx^i dx^j
\end{eqnarray}
again with $\bar{N}^2 > 0$ and $\bar{h}_{ij}$ a smooth Euclidean metric, then $R^H_{\mu\nu}$ shares the static symmetry.
Then due to this static symmetry the Harmonic Einstein equation $R^H_{\mu\nu} = 0$ thought of as PDEs for the metric components of $g$ is invariant under an analytic continuation $t \rightarrow  \tau = i t$. Hence we immediately see that the Harmonic Einstein equation restricted to Lorentzian \emph{static} metrics and reference metrics is elliptic. The behaviour of local relaxation or Ricci flow, and the Newton method will be precisely the same in either signature. 

Under a Euclidean continuation, with Euclidean time taken to be periodic, we could remove the boundary associated to a horizon. However, we noted that in practice one should take advantage of the static isometry and adapt coordinates to it, but then the horizon manifests itself as a fictitious  boundary of such coordinates, analogous to the origin of polar coordinates. Boundary conditions at the horizon in these adapted coordinates are just derived from transforming to regular coordinates which do not manifest the isometry but do manifest smoothness. In Lorentzian signature we have no option but to think of the horizon as a boundary. However, since the Harmonic Einstein equations are independent of signature, the boundary conditions for a regular Lorentzian horizon are precisely the same as those in Euclidean signature. Let us take coordinates in the base adapted to the horizon such that $x^i = (r , x^a)$ where $r =0 $ is the horizon. Then we write,
\begin{eqnarray}
\label{eq:staticpolarEuc}
ds^2 =  - r^2 V dt^2 + U  dr^2 + r \, U_a dr dx^a  + h_{ab} dx^a dx^b
\end{eqnarray}
where the metric functions are functions of $r$ and $x^a$. Changing to coordinates,
\begin{eqnarray}
a = r \cosh{\kappa t} \, , \quad b = r \sinh{\kappa t} 
\end{eqnarray}
provides a good chart covering the static Killing horizon, 
such that the metric components are smooth functions, provided that
$V, U, U_a, h_{ab}$ are smooth ($C^\infty$) functions of $r^2$ and $x^a$, and,
\begin{eqnarray}
V = \kappa^2  U
\end{eqnarray}
at the horizon $r =0$, and $\kappa$ is again the surface gravity.
The same conditions will apply to the reference metric which is also required to be smooth at $r = 0$. Of course we just have exactly the same situation as in the Euclidean case in equation \eqref{eq:Euchoriz} in \S\ref{sec:fictitious} with time continued back to Lorentzian signature. 

In the Euclidean context it is clear that since the metric is smooth and without boundary at the horizon, $R^H_{\mu\nu}$ must also be smooth there. The same is true in the Lorentzian case where the tensor $R^H_{\mu\nu}$ shares the same regularity properties as the metric. This is simply seen by noting that in the coordinates $(a, b, x^a)$ the metric and reference metric components are smooth functions, and hence so are those of $R^H_{\mu\nu}$. Transforming back to the static adapted coordinates $(t, r, x^a)$ then gives,
 \begin{eqnarray}
R^H =  - r^2 f dt^2 + g dr^2 + r \, g_a dr dx^a + r_{ab} dx^a dx^b
\end{eqnarray}
where $f, g, g_a$ and $r_{ab}$ are smooth in $r^2, x^a$, and in addition $f = \kappa^2 g$.   
Thus in the Lorentzian picture we have the nice property that Ricci flow and the Newton method will preserve the 
regularity of the horizon boundary. Furthermore it will preserve the surface gravity of the horizon.
Thus we must now view the horizon as a boundary and we are naturally lead to impose physical data there, namely the surface gravity with respect to $\partial / \partial t$.

\subsection{Stationary spacetimes with globally timelike Killing vector}
\label{sec:stationary}

We begin our discussion of stationary spacetimes by considering the case of spacetimes with globally timelike Killing vector, and will argue that the Harmonic Einstein equation is elliptic. Of course we are ultimately interested in black hole spacetimes which violate such a condition, with the stationary Killing vector becoming either null on the horizon, or outside the horizon at the boundary of an ergoregion.
In the following section \S\ref{sec:blackholes} we consider more general stationary spacetimes which allow horizons and ergoregions.

Consider the most general stationary metric with Killing vector $T = \partial / \partial t$, which we may write using coordinates adapted to the stationary isometry as,
\begin{eqnarray}
\label{eq:stationarygeneral}
ds^2 = - N(x) \left( dt + A_i(x) dx^i \right)^2 + h_{ij}(x) dx^i dx^j \, .
\end{eqnarray}
Now under our assumption that $T$ is globally timelike we have $N > 0$ and we further assume that the function $N$ is bounded. Physically this implies our spacetime has no Killing horizons, and also no ergoregions.
Since $\det{ g_{\mu\nu}} = - N \det h_{ij}$ we see that provided the metric $g$ is Lorentzian and smooth, so that $\det g_{\mu\nu} < 0$ and bounded,  this implies that $\det{h_{ij}} > 0$. We may then regard this metric as a smooth fibration of time over a base manifold $\mathcal{M}$ so that $(\mathcal{M},h)$ is a smooth Riemannian manifold with Euclidean signature metric $h_{ij}$. It is worth noting that this metric is not of the ADM form, but rather takes the form of a Kaluza-Klein reduction ansatz with respect to time. Thus $h_{ij}$ does not give the metric of a constant time slice of the Lorentzian geometry.

The second order derivative terms acting on the metric components $g_{\mu\nu}$ in the stationary Harmonic Einstein equation go as,
\begin{eqnarray}
{R}^H_{\mu\nu} = -\frac{1}{2} g^{\alpha\beta} \partial_ \alpha \partial_\beta {g}_{\mu\nu} + \ldots = -\frac{1}{2} h^{ij} \partial_i \partial_j {g}_{\mu\nu} + \ldots
\end{eqnarray}
where $\ldots$ are lower derivative terms. 
We see whilst the metric $g_{\mu\nu}$ is indeed Lorentzian, since there is no dependence on the coordinate $t$, it is actually the metric $h_{ij}$ that controls the character. We note the Kaluza-Klein form above ensures that the inverse metric $g^{\mu\nu}$ in the base directions is simply given in terms of the inverse of $h_{ij}$. 
This immediately implies that the Harmonic Einstein equation $R^H_{\mu\nu} = 0$ is elliptic since $h$ is smooth and of Euclidean signature. 

We also require that $R^H_{\mu\nu}$ is a tensor that is symmetric with respect to the stationary isometry $T$. Without this, Ricci-DeTurck flow and Newton's method will not consistently truncate to the class of stationary metrics \eqref{eq:stationarygeneral}. In order that $R^H_{\mu\nu}$ preserves the symmetry $T$, we choose the reference metric $\bar{g}$ to also be a smooth Lorentzian metric which is stationary with respect to the vector field $T$, so that,
\begin{eqnarray}
\bar{g} = - \bar{N}(x) \left( dt + \bar{A}_i(x) dx^i \right)^2 + \bar{h}_{ij}(x) dx^i dx^j 
\end{eqnarray}
where we also assume here that $T$ is globally timelike and bounded with respect to $\bar{g}$ so that $\bar{N} > 0$ and bounded. Then $\bar{h}_{ij}$ gives a second Riemannian metric on the same manifold $\mathcal{M}$. Since now $R^H_{\mu\nu}$ preserves the stationary symmetry, the Ricci-DeTurck flow can be consistently truncated to a parabolic flow on the space of Lorentzian stationary metrics. Since this flow remains diffeomorphic to Ricci flow (subject at least to the normal component of $\xi$ vanishing on any boundaries), we arrive at the interesting result that we may apply parabolic Ricci flow to \emph{stationary Lorentzian} spacetimes. Likewise the Newton method will preserve the stationary symmetry and can be used to solve this elliptic stationary problem.

In a situation where the solution we wish to find has a stationary Killing vector that is globally timelike and bounded then nearby to that solution the character of the Harmonic Einstein equations will be elliptic. Subject to imposing suitable boundary conditions on any boundaries or asymptotic regions, one may use the Lorentzian stationary Ricci-DeTurck flow or Newton method to solve for the solution. One must start with an initial guess that has globally timelike bounded stationary Killing field $T$, and then provided that guess is sufficiently good, one can hope the subsequent Ricci-DeTurck flow or Newton iterations preserve that $T$ is globally timelike.

\subsection{Stationary black holes}
\label{sec:blackholes}

We now proceed to consider the case of non-extremal black holes.
 In the context of the discussion above now the norm of $T$ will vanish either at the horizon itself, assuming that $T$ is a globally timelike Killing vector (such as exist for certain Kerr-AdS black holes), or outside the horizon in the ergoregion. Since we are interested in the exterior of the horizon, in the first case we may treat the system described above for globally timelike  $T$ and now regard the horizon as a boundary  where suitable boundary conditions are required. 
However in the latter, more general case, outside the horizon but inside the ergoregion we have the norm of $T > 0$ and hence $\det{h_{ij}} < 0$. Now the base manifold in the previous section fails to be Riemannian and then our argument above that the Harmonic Einstein equation is elliptic fails. 

In order to make progress we must use the Rigidity property of stationary black holes, proved in $D> 4$ by Ishibashi, Hollands and Wald \cite{Hollands:2006rj} for 
various asymptotics, including asymptotically flat solutions.
Assume there exists a stationary Killing vector $T$. Then the Rigidity theorem states that for an asymptotically flat rotating black hole, so that $T$ is not normal to the horizon, there exists a Killing vector $K$ that commutes with $T$ and which is normal to the horizon. Furthermore there exist some number $N \ge 1$ of commuting Killing vectors $R_a$, which also commute with $T$ and asymptotically generate spatial rotation with closed orbits of period $2 \pi$. The theorem states that $K$ may be written in terms of these as, $K = T + \Omega^a R_a$,
for some constants $\Omega^a$.  Consequently the horizon rigidly moves with respect to the orbits of $K$ in its exterior, and hence with respect to the asymptotic rotation generators $R_a$.
Were this not the case one would expect gravitational radiation to be emitted from the region near the horizon and this would presumably violate the assumption of stationarity.

Let us proceed by assuming Rigidity holds so that there exists a stationary Killing vector $T$ and Killing vectors $R_a$ for $a = 1,\ldots,N$, and the vector fields $T$ and $R_a$ are all commuting. We take the vectors $R_a$ to generate spatial isometries with either compact or non-compact orbits. In the compact case we take the period to be $2 \pi$, and we allow axes of this symmetry where the isometry has fixed action. Rigidity implies we may write the normal $K$ to our Killing horizon as,
\begin{eqnarray}
K = T + \Omega^a R_a \, .
\end{eqnarray}
We may write the metric adapting coordinates to the isometries,  
\begin{eqnarray}
\label{eq:bhansatz}
d{s}^2  = {G}_{AB}(x) \left( dy^A + {A}^A_i(x) dx^i \right) \left( dy^B +{A}^B_j(x) dx^j \right) +{h}_{ij}(x) dx^i dx^j
\end{eqnarray}
where $y^A = \{ t, y^a \}$ and $T = \partial / \partial t$ and $R_a = \partial / \partial y^a$. In analogy with the stationary case in the previous section we see that the geometry may be thought of as a fibration of the Killing vector directions over a base manifold $\mathcal{M}$ with metric $h_{ij}$. Technically $\mathcal{M}$ is the orbit space of the full Lorentzian spacetime with respect to the isometries $T, R_a$.
We note that whilst in 4D for vacuum asymptotically flat solutions the circularity theorem implies one can find a coordinate system where the cross terms between base and fibre, $A^A_i$, vanish this is not expected to be the case for general stationary black holes in higher dimensions. At present, however, the only known solutions do in fact have vanishing $A^A_i$.

As with the analytic work on uniqueness, the aim now is to formulate the problem as an elliptic one on the orbit space $\mathcal{M}$.
We emphasize that here we are trying to find constructive numerical techniques to find black holes, rather than to prove their existence or uniqueness. With this in mind we make our key assumption;\\

{\bf Assumption:} $(\mathcal{M},h)$ is a smooth Riemannian manifold.\\

The full spacetime is Lorentzian, and so exterior to the horizon $\det g_{\mu\nu} = \det G_{AB} \det h_{ij} < 0$. The chart breaks down at the horizon where $\det G_{AB} = 0$ since the norm of $K$ vanishes. It also breaks down at an axis of symmetry where some $R_a$ vanishes and again $\det G_{AB} = 0$. However, our assumption ensures that to the exterior of all horizons and axes then $\det G_{AB} < 0$ and hence $G_{AB}$ is of Lorentzian signature. We regard the horizon and axes of symmetry of the $R_a$'s as boundaries for the base manifold $\mathcal{M}$. We note our assumption above ensures that the geometry of these boundaries is smooth. 
For simplicity we assume here that the boundaries are only due to the horizon and vanishing of various $R_a$'s. However more generally one might consider multiple Killing horizons, and boundaries where linear combinations of the $R_a$'s vanish.\footnote{This is discussed in detail in the case of $D-2$ commuting Killing vectors \cite{Hollands:2007aj}.}

Harmark has discussed the above form of metric in the context of classifying stationary spacetimes \cite{Harmark:2009dh}. The structure of $\mathcal{M}$ together with the data $\Omega^a$ at the horizon (or more generally horizons), and the data of which combination of $R_a$'s vanishes at the axis boundaries defines a `rod structure' for stationary spacetimes and has been conjectured to classify higher dimensional black holes.

It is instructive to consider the simple example of the Kerr solution from this perspective of the time and rotation Killing directions being fibred over a smooth base.
 In the conventional Boyer-Lindquist coordinates the Kerr metric takes the form,
\begin{eqnarray}
ds^2 = G_{tt} dt^2 + 2 G_{t\phi} dt d\phi + G_{\phi\phi} d\phi^2 + h_{rr} dr^2 + h_{\theta\theta} d\theta^2
\end{eqnarray}
with vanishing $A^A_i$ where,
\begin{eqnarray}
&&G_{tt} = - \frac{ \left(  \Delta - a^2 \sin^2 \theta \right) }{ \Sigma }  \, , \quad
G_{\phi\phi} =   \sin^2 \theta  \frac{ \left( ( r^2 + a^2 )^2 - \Delta a^2 \sin^2\theta \right) }{ \Sigma} \, , \nonumber \\
&&G_{t\phi} = - a \sin^2\theta \frac{ \left( r^2 + a^2 - \Delta \right) }{ \Sigma}  \, , \quad h_{rr} = \frac{\Sigma}{\Delta}  \, , \quad h_{\theta\theta} = \Sigma
\end{eqnarray}
with $\Delta = r^2 + a^2 - 2 M r$ and $\Sigma = r^2 + a^2 \cos^2\theta$. Here $ T = \pd{}{t}$ and $R = \pd{}{\phi}$. The outer horizon and axis are the boundaries of the base manifold $\mathcal{M}$ and are located at $r = r_h$ (where $\Delta = 0$) and $\theta = 0$, $\pi$ respectively. The Killing field $K = T + \Omega R$ is tangent to the horizon and timelike near there, where the angular velocity of the horizon is given as, $\Omega = \frac{a}{a^2+ r_h^2}$. 
One finds,
$\det{G_{AB}} = - \Delta$,
which vanishes at the horizon and axis, but not in their exterior. Whilst the $\theta$ coordinate is a regular coordinate on the base at the rotation axes, the radial $r$ coordinate is not at the horizon since $\Delta$ vanishes and so $h_{rr} \to \infty$ there. We therefore define a new radial coordinate, $\rho$, such that $d \rho = dr / \sqrt{\Delta}$ and $\rho = 0$ at the horizon, giving
$r = M + \sqrt{M^2 - a^2} \cosh \rho$,
so that the components of the base metric $h_{ij}$ are smooth at the horizon boundary. In particular in these coordinates the determinant of the base metric, 
\begin{eqnarray}
h_{ij} dx^i dx^j = \frac{\Sigma}{\Delta} dr^2 + \Sigma d\theta^2 = \Sigma \left( d\rho^2 + d\theta^2 \right)  \quad \implies \quad \det{h_{ij}} = \Sigma^2 \ge  r_h^2
\end{eqnarray}
and thus we see that since $r_h > 0$ the base is indeed a smooth Riemannian manifold everywhere on and in the exterior of the horizon and axis of symmetry.

\subsection{Ellipticity of the stationary problem}

We note that we have not required the stationary Killing field $T$ to be timelike. In the presence of horizons it will become null on the horizon or be spacelike if the horizon is surrounded by an ergoregion. We reiterate that in the previous section \S\ref{sec:stationary} it was precisely where $T$ failed to be timelike that ellipticity would break down, since the base metric would fail to be Riemannian. The crucial observation is that for our class of stationary spacetimes \eqref{eq:bhansatz},
\begin{eqnarray}
R^H_{AB} & = & -\frac{1}{2} g^{\alpha\beta} \partial_ \alpha \partial_\beta {g}_{AB} + \ldots = -\frac{1}{2} h^{mn} \partial_m\partial_n {G}_{AB} + \ldots \nonumber \\
R^H_{Ai} & = & -\frac{1}{2} g^{\alpha\beta} \partial_ \alpha \partial_\beta {g}_{Ai} + \ldots = -\frac{1}{2} h^{mn} \partial_m\partial_n \left( G_{AB} A^B_{i} \right) + \ldots \\
R^H_{ij} & = & -\frac{1}{2} g^{\alpha\beta} \partial_ \alpha \partial_\beta {g}_{ij} + \ldots = -\frac{1}{2} h^{mn} \partial_m \partial_n \left( {h}_{ij} + G_{AB} A^A_{i}A^B_{j}\right)+ \ldots \nonumber
\end{eqnarray}
where again the $\ldots$ represent lower than second order derivative terms. We see the equations have character determined solely by the metric $h_{ij}$, and by our assumption above that the base $\mathcal{M}$ is Riemannian,  this is indeed elliptic. Ergo-regions may occur where $T$ is no longer timelike, but our assumption that the base is Riemannian implies that some linear combination of the Killing directions $T, R_a$ is always timelike outside the horizons. 

In analogy with the previous section \ref{sec:stationary}, in order to ensure that $R^H_{\mu\nu}$ shares the symmetries of $g$  we choose the reference metric $\bar{g}$ so that $T, R_a$ are again Killing with respect to it, and obey precisely the same assumptions as above for $g$. Thus we may write,
\begin{eqnarray}
\label{eq:bhref}
\bar{ds}^2 &=& \bar{g}_{\mu\nu} dX^\mu dX^\nu \\
&&= \bar{G}_{AB}(x) \left( dy^A + \bar{A}^A_i(x) dx^i \right) \left( dy^B + \bar{A}^B_j(x) dx^j \right) + \bar{h}_{ij}(x) dx^i dx^j \nonumber
\end{eqnarray}
and we further assume that $(\mathcal{M},\bar{h})$ is a smooth Riemannian manifold. Then the Ricci-DeTurck flow and Newton's method consistently truncate to the Lorentzian stationary spacetimes of the form \eqref{eq:bhansatz}.

We must impose suitable boundary conditions at the boundaries of $\mathcal{M}$. Asymptotically we might impose the asymptotic flatness or Kaluza-Klein conditions mentioned above, which are compatible with $\xi \to 0$. The new feature is that we have additional boundaries on the base corresponding to the Killing horizons and axes of symmetry and we will discuss this shortly. 
Using the Ricci-DeTurck flow or Newton method if we start from initial data in our stationary class, then for small flow times or updates we expect to remain in this class. In particular we expect $(\mathcal{M},h)$ to remain a Riemannian manifold. Provided this condition holds for the solution of interest, and our initial guess is sufficiently close to this, then we might hope to reach this solution. 

We note that the maximum principle discussed in the static context relied on the inequality in \eqref{eq:phicondition}, which results from the positivity of $(\nabla_{\mu} \xi_\nu) (\nabla^\mu \xi^\nu) \ge 0$ for a Riemannian manifold. Since the equations for the static case are independent of signature the maximum principle must equally apply in the static Lorentzian context. However, one can check that in the stationary case this term has indefinite sign, and it is unclear if a maximum principle can be found ruling out solitons. This is not a problem in practice where one must simply check whether a solution obtained is a soliton or not. However, the elegant property that in certain static cases there can be no solitons does not obviously generalise to the stationary case.

\subsection{Boundary conditions for the stationary problem}
\label{sec:bc}

We conclude this chapter by now explicitly giving the boundary conditions for the metric components of our stationary spacetime \eqref{eq:bhansatz} at the Killing horizon or symmetry axes. 
These are a generalisation of the boundary conditions determined for the 4D stationary axisymmetric vacuum problem of the classic uniqueness theorems \cite{Carter}.
They are derived and discussed in more detail in \cite{newpaper} and are consistent with the boundary conditions discussed by Harmark using particular coordinates on the base manifold \cite{Harmark:2009dh}.
Consider a Killing horizon with $K = T + \Omega^a R_a$. It is then convenient to change coordinates as,
\begin{eqnarray}
\label{eq:trans}
t \, , \; y^a \quad \rightarrow \quad \tilde{t} = t \, , \quad \tilde{y}^a = y^a - \Omega^a t
\end{eqnarray}
so that $K = \partial / \partial \tilde{t}$ and $R_a = \partial / \partial \tilde{y}^a$. Note that if $y^a$ is periodic then $\tilde{y}^a$ is also a periodic coordinate with period $2 \pi$.
Now consider a boundary, either due to the vanishing of $K$ or a compact $R_a$. We take base coordinates $x^i = ( r , x^{\tilde{i}} )$ adapted to the boundary so that it lies at $r = 0$, and decompose the base metric as,
\begin{eqnarray}
\label{eq:hdecomp}
h_{ij} dx^i dx^j = N dr^2 + r \, N_{\tilde{i}} d r dx^{\tilde{i}} + h_{\tilde{i}\tilde{j}} dx^{\tilde{i}} dx^{\tilde{j}}
\end{eqnarray}

\noindent {\bf Horizon:}  For a Killing horizon we write the following metric components as,
\begin{eqnarray}
G_{\tilde{t} A} = - r^2  f_A \, , \quad A^{A}_{r} = r g^A \, , 
\end{eqnarray}
for $A = (\tilde{t}, \tilde{y}^a)$ and then let $X = \left\{ f_A \, , \;  g^A \, , \; G_{\tilde{y}^a \tilde{y}^b}  \, , \; 
  A^{A}_{\tilde{i}}\, , \;  
N \, , \; N_{\tilde{i}} \, , \; h_{\tilde{i}\tilde{j}} \right\}
$ be the set of functions describing our metric. 
Then by considering a change of coordinates,
\begin{eqnarray}
a = r \cosh{\kappa \tilde t} \, , \quad b = r \sinh{\kappa \tilde t} 
\end{eqnarray}
and requiring in these Cartesian coordinates the components are smooth, we deduce the following behaviour is required in the stationary adapted chart; the functions $X$ must be smooth functions of $r^2$ and $x^{\tilde{i}}$ at $r=0$, and furthermore  obey a regularity condition,
\begin{eqnarray}
\left( f_{\tilde{t}} - \kappa^2 N \right) |_{r=0} = 0
\end{eqnarray}
where $\kappa$ is constant and gives the surface gravity with respect to $K$.\\

\noindent {\bf Axis:}
Consider an axis associated to a vanishing compact $R_a$. Without loss of generality choose this to be $R_N$. Then we choose to write,
\begin{eqnarray}
G_{\tilde{y}^N A} = r^2  f_A \, , \quad A^{A}_{r} = r g^A \, , 
\end{eqnarray}
and let $Y = \left\{ f_A \, , \;  g^A \, , \; G_{\tilde{t}\tilde{t}}  \, , \;  G_{\tilde{t}\tilde{y}^{\tilde{a}}}  \, , \;  G_{\tilde{y}^{\tilde{a}}\tilde{y}^{\tilde{b}}}  \, , \;  A^{A}_{\tilde{i}}\, , \;  
N \, , \; N_{\tilde{i}} \, , \; h_{\tilde{i}\tilde{j}} \right\}
$  be the set of functions describing our metric (where $\tilde{a} = 1,\ldots,N-1$). A similar analysis as for the shrinking of the Euclidean time circle previously 
 implies that for a smooth metric we must have that the metric functions $Y$ are smooth functions of $r^2$ and $x^{\tilde{i}}$ at $r=0$, and in addition we require,
\begin{eqnarray}
\label{eq:axis}
\left( f_{\tilde{y}^N} - N \right) |_{r=0} = 0
\end{eqnarray}
Of course we obtain analogous conditions for an axis with respect to a different $R_a$. \\

It is straightforward to check that the boundary conditions at the meeting of a horizon with an axis, or two axes, are compatible with each other. Take coordinates in the base $x^i = ( r_1 , r_2, x^{\tilde{i}} )$ where $r_1 = 0$ gives the position of the first boundary, and $r_2 = 0$ gives the second boundary, and hence the origin $r_1 = r_2 = 0$ is the meeting point. The boundary conditions near this origin are simply the union of the boundary conditions for each boundary. Note this implies that two boundaries (a horizon and axis, or two axes) meet in the base at right-angles. We reiterate that we have only considered axes arising from fixed points of the $R_a$'s, and more generally one could consider linear combinations of these vanishing.

A very important point is that as discussed in the earlier section \ref{sec:harmeqn} having introduced boundary conditions we must check that these are compatible with finding Ricci flat solutions. To investigate this we must consider our choice of reference metric \eqref{eq:bhref}, which also 
 is required to be regular and hence is subject to the same  boundary conditions above for its components on the various horizon and axis boundaries. In particular we note that the surface gravity of the reference metric horizon must be the same as that of the actual metric. 
 One can then explicitly check that,
\begin{eqnarray}
\xi^r |_{r = 0} = 0 \, , \quad \partial_r \xi^{\tilde{i}} |_{r = 0} = 0 \, , \quad \partial_r \xi^A |_{r = 0}= 0
\end{eqnarray}
both at a horizon and axis of symmetry, which is indeed consistent with the linear elliptic problem $\mathcal{D} \cdot \chi = 0$ discussed in section \ref{sec:solitons} being well posed and admitting the trivial solution.
Note that since $\xi^r = 0$ the Ricci-DeTurck flow should be diffeomorphic to Ricci flow in the presence of such boundaries. 
Furthermore the Harmonic Einstein tensor will be regular at the horizon and axis boundaries. Thus in our adapted coordinates it will also obey the same regularity conditions as the metric above. In particular, Ricci-DeTurck flow and Newton's method will preserve regularity, and will have the elegant result that they will leave the surface gravity constant. \\

\vspace{0.5cm}
\noindent
{ \it This work is dedicated to my father. }

\section*{Acknowledgements}

I am greatly indebted to my collaborators Alexander Adam, Pau Figueras, Matthew Headrick, James Lucietti and Sam Kitchen. 

%\bibliography{refnum}
%\bibliographystyle{unsrt}

\end{document}